\documentstyle[12pt]{article}
\def\lsim{\mathrel{\rlap {\raise.5ex\hbox{$ < $}}
{\lower.5ex\hbox{$\sim$}}}}
\def\gsim{\mathrel{\rlap {\raise.5ex\hbox{$ > $}}
{\lower.5ex\hbox{$\sim$}}}}

\newcommand{\pr}{\paragraph{}}
\newcommand{\be}{\begin{equation}}
\newcommand{\ee}{\end{equation}}
\newcommand{\bea}{\begin{eqnarray}}
\newcommand{\nn}{\nonumber}
\newcommand{\eea}{\end{eqnarray}}

\newcommand{\nk}{\noindent}
\newcommand{\BbbR}{I \! \!  R}

\def\gappeq{\mathrel{\rlap {\raise.5ex\hbox{$>$}}
{\lower.5ex\hbox{$\sim$}}}}

\def\lappeq{\mathrel{\rlap{\raise.5ex\hbox{$<$}}
{\lower.5ex\hbox{$\sim$}}}}

\begin{document}

\begin{titlepage}
\begin{flushright}
OUTP-95-37P, hep-th/9510007 \\
\end{flushright}

\begin{centering}
\vspace{.1in}
{\large {\bf Aspects of hairy black holes in spontaneously-broken
Einstein-Yang-Mills systems: Stability analysis and Entropy considerations
 }} \\
\vspace{.2in}
{\bf N.E. Mavromatos}$^{*}$ and {\bf Elizabeth Winstanley} \\
\vspace{.1in}
\vspace{.03in}
Department of Physics
(Theoretical Physics), University of Oxford, 1 Keble Road,
Oxford OX1 3NP, U.K.  \\

\vspace{.1in}
{\bf Abstract} \\
\vspace{.05in}
\end{centering}
{ We analyze (3+1)-dimensional black-hole space-times in
spontaneously
broken Yang-Mills gauge
theories that have been
recently presented as candidates
for an evasion of the scalar-no-hair theorem.
Although we show that in principle
the conditions for the no-hair theorem do not apply
to this case, however we prove that the `spirit' of the theorem
is not violated, in the sense that
there exist instabilities, in both the sphaleron and
gravitational sectors.
The instability analysis of the sphaleron sector, which was expected
to be unstable for topological reasons,
is performed
by means of a variational method. As shown,
there exist modes in this sector
that are unstable against linear perturbations.
Instabilities exist also in the gravitational sector.
A method for counting
the gravitational unstable modes, which utilizes
a catastrophe-theoretic approach is presented.
The r\^ole of the catastrophe functional
is played by the mass functional of the black hole.
The Higgs vacuum expectation value (v.e.v.)
is used as a control
parameter, having a critical value beyond which
instabilities are turned on. The (stable)
Schwarzschild solution is then understood from this point of view.
The catastrophe-theory appproach
facilitates
enormously a universal stability study of non-Abelian black holes,
which goes beyond linearized perturbations.
Some elementary entropy considerations are also
presented that support the catastrophe theory analysis, in the
sense that `high-entropy' branches of solutions
are shown to be relatively more stable than `low-entropy' ones.
As a partial result of this entropy analysis, it is also shown
that there exist {\it logarithmic} divergencies
in the entropy of matter (scalar) fields near the horizon, which are
up and above the linear divergencies, and, unlike them, they
cannot be absorbed in a
renormalization of the gravitational coupling constant
of the theory. The associated part of the entropy
violates the classical Bekenstein-Hawking formula which is a proportionality
relation between black-hole entropy and horizon area.
Such logarithmic divergencies, which are associated with the
presence of non-abelian gauge and Higgs fields,
persist in the `extreme case',
where linear divergencies disappear.
}
\vspace{0.1in}
%
\pr

\vspace{0.01in}
\begin{flushleft}
$^{*}$P.P.A.R.C. Advanced Fellow \\
September 1995 \\
\end{flushleft}
\end{titlepage}
\newpage
\section{Introduction}
\pr
The surprising discovery of the Bartnik-McKinnon (BM) non-trivial particle-like
structure~\cite{bartnik} in the Einstein-Yang-Mills system
opened many possibilities for the existence of non-trivial
solutions to Einstein-non-Abelian-gauge systems.
Indeed, soon after its discovery, many other
self-gravitating structures with non-Abelian gauge fields
have been discovered~\cite{selfgrav}. These
include black holes
with non-trivial hair, thereby leading to the possibility of
evading the no-hair conjecture~\cite{nohair}.
The physical reason for the existence of these classical
solutions is the `balance' between
the non-Abelian gauge-field repulsion
and the gravitational attraction. Such a balance
allows for dressing black hole solutions
by non-trivial configurations (outside the horizon)
of fields that are not associated with a Gauss-law,
thereby leading to an `apparent' evasion of the no-hair conjecture.
\pr
Among such black-hole solutions, a physically interesting
case is that of a spontaneously broken Yang-Mills theory
in a non-trivial black-hole space-time (EYMH)~\cite{greene}.
This system has been recently examined from a stability point of view,
and found to possess an instability~\cite{winst},
thereby making the physical importance
of the solution rather marginal, but also indicating another
dimension of the no-hair conjecture, not considered in the original
analysis, that of stability.
\pr
In this article, we shall give more details of this stability considerations
by extending the analysis to incorporate counting of the
unstable modes, and going beyond the linear-stability case
by employing catastrophe theory~\cite{torii}
in order to analyse
instabilities
in the gravitational sector of the solution.
Catastrophe theory is a powerful mathematical tool
to study or explain a variety of change of states in
nature, and in particular a {\it discontinuous }
change of states that occurs eventually despite a
gradual (smooth) change of certain parameters of the
system. In the case at hand, the catastrophe
functional,
which exhibits a discontinuous change in its
behaviour, will be the mass of the black hole space time,
whilst the control parameter, whose smooth change
turns on the catastrophe at a given critical value, will
be the vacuum expectation value (v.e.v.) of the Higgs field.
The advantage of using the v.e.v. of the Higgs field as the control
parameter, rather than the horizon radius as was done in \cite{torii},
is that it will allow us to relate the stability of the
EYMH black holes  to that of the Schwarzschild solution, which
is well known to be stable.
The type of catastrophe encountered will be that of a {\it fold}
catastrophe.
The catastrophe-theoretic approach allows for a
universal stability study of non-abelian black hole solutions
that goes beyond linearised perturbations; the particular
use of the Higgs v.e.v. as a control parameter
in the case of the EYMH systems allows an
{\it exact} counting of the unstable modes.
\pr
As part of our analysis, we shall make an attempt to associate
the above catastrophe-theoretic considerations with
some `thermodynamic/information-theoretic' aspects of black hole physics,
and in particular with the entropy of the black hole.
By computing explicitly the entropy of
quantum fluctuations of (scalar) matter fields
near the horizon we shall show that `high-entropy' branches
of the solution possess less unstable modes (in the gravitational sector)
than the `low-entropy' ones. As a partial, but not less important, result
of this part of our analysis, we shall also show that
the entropy of the black hole possess linear {\it and} logarithmic
divergencies. The linear divergencies do not violate the Bekenstein-Hawking
formula relating entropy to the  classical horizon area. The only
difference is the divergent proportionality factors in front, which, however,
can be absorbed in a conjectured renormalization of Newton's constant in the
model~\cite{thooft,susskind}. This is not the case with the logarithmic
divergencies
though. The latter persist even in `extreme black-hole' cases,
where the linear divergencies disappear.
They clearly violate the Bekenstein-Hawking
formula. In our case they owe
their presence to the non-Abelian gauge and Higgs fields.
The presence of logarithmic divergencies in black hole physics
has been noted in ref. \cite{susskind}, but only in examples
involving truncation from (3+1)-dimensional space-times to (1 + 1) dimensions,
and
in that reference their
presence
had been attributed to this bad truncation of the four-dimensional
black hole spectrum. Later on, however, such logarithmic divergencies
have been confirmed to exist in
string-inspired dilatonic black holes in (3+1)
dimensions~\cite{dilatonbh}. Their presence in our EYMH system,
and in general in non-Abelian black holes as we shall show, indicates
that such
logarithmic divergencies are {\it generic}
in black hole space-times with non-conventional hair, and probably indicates
information loss, even in extreme cases, associated with the
presence of space time boundaries.
This probably implies
that the entropy of the black hole is not only associated with
classical geometric factors, but
is a much more complicated phenomenon related
to information carried by the various (internal) black hole states.
The latter
phenomenon could be associated with, and may be offer ways out of,
the usual difficulties of reconciling quantum mechanics
with canonical quantum gravity.
\pr
The structure of the article
is as follows. In section 2 we shall discuss the no hair conjecture
for black holes space-times with non-trivial scalar field
configurations by following
a modern approach due to Bekenstein~\cite{bekmod}.
We shall show that the proof of the no-hair theorem fails for the
case of the EYMH system, in accordance with the explicit
solution found in ref. \cite{greene}. In section 3 we shall present
a stability analysis of the system based on linear perturbations.
We shall demonstrate the existence of instabilities
in the sphaleron sector, following a variational
approach which is an extension of the approach of
Volkov and Gal'tsov~\cite{volkov}
to study particle-like solutions. We shall also
present arguments for counting the unstable modes in the sphaleron
sector of the theory.
In section 4 we shall present a method for counting the unstable
modes in the gravitational sector by
going beyond  the linearised-perturbation analysis using catastrophe
theory, with the mass functional of the black hole as the
catastrophe functional and the Higgs v.e.v. as the control parameter.
In section 5, in
connection with the latter approach,  we shall estimate
the entropy of the various branches of the solution using a WKB approximation.
We shall show that the high-entropy branch of solutions
is relatively more stable (in the sense of possesssing fewer unstable
modes) than the low-entropy branch.
As we have already mentioned, we shall also discuss the existence
of logarithmic divergencies
in the entropy, associated with the presence of
non-trivial hair in the black hole, in certain extreme cases, and we shall
argue about an explicit violation of the classical Bekenstein-Hawking
entropy formula, indicating a different
(information oriented) r\^ole of the black hole entropy.
Conclusions and outlook will be presented in section 6.
Some technical aspects of our approach will be discussed in two appendices.

\section{Bypassing Bekenstein's no-hair theorem in
\newline EYMH systems}
\pr
Recently, Bekenstein presented a modern elegant proof
of the no-hair theorem for black holes, which covers
a variety of cases with scalar fields~\cite{bekmod}. The theorem
is formulated in such a way so as to rule out a multicomponent
scalar field dressing an asymptotically flat, static, spherically-symmetric
black hole. The basic assumption of the theorem is that the scalar
field is minimally coupled to gravity and bears a non-negative energy density
as seen by any observer, and the proof relies on very general principles,
such as energy-momentum conservation and the Einstein equations.
{}From the positivity assumption and the conservation equations
for the energy momentum tensor $T_{MN}$ of the theory, $\nabla ^M T_{MN}= 0$,
one obtains for a spherically-symmetric space-time background
the condition that near the horizon the radial component
of the energy-momentum tensor and its first derivative are negative
\be
    T _r^r < 0,  \qquad (T_r^r)' < 0
\label{zero}
\ee
with the prime denoting differentiation with respect to $r$.
This
implies that in such systems there must be regions in space, outside
the horizon where both quantities in (\ref{zero}) change sign.
This contradicts the results following from Einstein's equations
though~\cite{bekmod}, and this {\it contradiction} constitutes
the proof of the no-hair theorem, since the only allowed non-trivial
configurations are Schwarzschild black holes.
We note, in passing,
that there are known exceptions to the original version of the
no-hair theorem~\cite{nohair}, such as conformal
scalar fields coupled to gravity, which
come from the fact that in such theories the scalar fields
diverge at the horizon of the black hole~\cite{confdiv}.
\pr
The interest for our case
is that the theorem rules out the existence
of non-trivial hair due to a Higgs field with a double (or multiple) well
potential, as is the case for spontaneous symmetry breaking.
Given that stability issues are not involved in the proof,
it is of interest to reconcile the results of the theorem with the situation
in our case of EYMH systems,
where at least
we know that an explicit solution with non-trivial
hair exists~\cite{greene}, albeit unstable~\cite{winst}.
As we shall
show below, the formal reason
for bypassing
the modern version
of the no-hair theorem~\cite{bekmod}
lies in the violation
of the key relation among the
components of the stress tensor, $T_t^t =T_\theta^\theta$,
shown to hold in the case of ref. \cite{bekmod}.
The physical reason
for the `elusion' of the above no-hair conjecture lies in the fact that
the presence of the repulsive non-Abelian gauge interactions
balance the gravitational attraction, by producing terms that
make the result (\ref{zero}) inapplicable in the present case.
Below we shall demonstrate this in a mathematically rigorous way.
\pr
To this end, consider the EYMH theory with Lagrangian
\be
{\cal L}_{EYMH} = -\frac{1}{4\pi} \left\{ \frac{1}{4} |F_{MN}|^2
+ \frac{1}{8} \phi ^2 |A_M|^2 + \frac{1}{2} |\partial _M \phi |^2
+ V(\phi ) \right\}
\label{onea}
\ee
where $A_M$ denotes the Yang-Mills field, $F_{MN}$ its field strength,
$\phi $ is the Higgs field and $V (\phi )$ its potential.
All the indices are contracted with the help of the
background gravitational tensor $g_{MN}$.
In the spirit of Bekenstein's
modern version of the no-hair theorem, we now examine
the energy-momentum tensor of the model (\ref{onea}). It can be written in the
form
\be
 8 \pi T_{MN} = -{\cal E } g_{MN} + \frac{1}{4\pi} \left\{ F_{MP}F_{N}{}^{P}
+ \frac{\phi ^2}{4} A_M A_N + \partial _M \phi \partial _N \phi \right\}
\label{threea}
\ee
with ${\cal E} \equiv -{\cal L}_{EYMH} $.
\pr
Consider, now, an observer moving with a four-velocity $u^M $ . The observer
sees a local energy density
\be
\rho  = {\cal E} + \frac{1}{4\pi} \left\{ u^M F_{MP} F_N{}^{P} u^N +
\frac{\phi ^2}{4} (u^M A_M )^2 + (u^M \partial _M \phi )^2 \right\},
\qquad u^M u_M = -1.
\label{foura}
\ee

To simplify the situation let us
consider a space-time with a time-like killing vector, and suppose that the
observer moves along this killing vector. Then
$ u^M \partial _M \phi = 0 $ and by an appropriate gauge choice
$u^M A_M =0 = u^M F_{MN} $. This gauge choice is compatible with the
spherically-symmetric ansatz
for $A_M$ of ref. \cite{greene}. Under these assumptions,
\be
    \rho = {\cal E}
\label{fivea}
\ee
and the requirement that the local energy density as seen by any observer
is non-negative
implies
\be
{\cal E } \ge 0.
\label{sixa}
\ee
\pr
We are now in position to proceed with the announced proof of the bypassing
of the no-hair theorem of ref. \cite{bekmod} for the EYMH black hole of
ref. \cite{greene}.
To this end we consider a spherically-symmetric ansatz for the
space-time metric $g_{MN}$, with an invariant line element of the form
\be
  ds^2 = - e^{\Gamma } dt^2 + e^\Lambda dr^2
+ r^2 (d\theta ^2 + \sin^2 \theta d\varphi ^2),
\qquad \Gamma = \Gamma (r),~\Lambda = \Lambda (r).
\label{sevena}
\ee
To make the connection with the black hole case we further assume that the
space-time is asymptotically flat.
\pr
{}From the conservation of the energy-momentum tensor, following from the
invariance of the effective action under general co-ordinate transformations,
one has  for the $r$-component of the conservation equation
\be
        [(-g)^{\frac{1}{2}} T_r^r ]' - \frac{1}{2} (-g)^{\frac{1}{2}}
\left( \frac{\partial }{\partial r} g_{MN}\right) T^{MN} = 0
\label{eighta}
\ee
with the prime denoting differentiation with respect to $r$. The spherical
symmetry of the space time implies that $T_\theta ^\theta =
T_\varphi ^\varphi $.
Hence, (\ref{eighta}) can be written as
\be
   (e^{\frac{\Gamma+\Lambda}{2}} r^2 T_r^r )' - \frac{1}{2}
e^{\frac{\Gamma + \Lambda}{2}} r^2 \left[ \Gamma' T_t^t + \Lambda ' T_r^r
+ \frac{4}{r}
T_\theta ^\theta \right] = 0.
\label{ninea}
\ee
Observing that the terms containing $\Lambda $ cancel,
and integrating over the radial coordinate $r$ from the horizon
$r_h$ to a generic distance $r$, one obtains
\be
   T_r^r (r) = \frac{e^{-\frac{\Gamma}{2}}}{2 r^2}
\int _{r_h}^r dr e^{\frac{\Gamma}{2}} r^2 \left[ \Gamma' T_t^t +
\frac{4}{r} T_\theta ^\theta \right]
\label{tena}
\ee
Note that
the assumption that scalar invariants, such as
$T_{MN}T^{MN}$  are finite on the horizon (in order that the latter
is regular),
implies that the boundary terms on the horizon vanish in
(\ref{tena}).
\pr
It is then straightforward to obtain
\be
     (T_r^r)' =\frac{1}{2} \left[ \Gamma' T_t^t
+ \frac{4}{r} T_\theta ^\theta \right]
- \frac{e^{-\frac{\Gamma}{2}}}{r^2} (e^{\frac{\Gamma}{2}} r^2)' T_r^r.
\label{elevena}
\ee
\pr
Next, we consider Yang-Mills fields of the form~\cite{greene}
\be
 A = (1 + \omega (r) ) [-{\hat \tau} _\phi d\theta + {\hat \tau}_\theta
\sin \theta d\varphi ]
\label{twelvea}
\ee
where $\tau _i $, $i =r,\theta,\varphi $
are the generators of the $SU(2)$ group
in spherical-polar coordinates.
Ansatz (\ref{twelvea}) yields
\bea
         T_t^t &=& - {\cal E} \nn \\
T_r^r &=& - {\cal E} + {\cal F}  \nn \\
T_\theta ^\theta & = & -{\cal E} + {\cal J}
\label{thirteena}
\eea
with (see Appendix A for details of the relevant quantities),
\bea
{\cal F} &\equiv & \frac{e^{-\Lambda}}{4\pi} \left[ \frac{2\omega '^2}{r^2}
+ \phi '^2 \right]
\nn  \\
{\cal J} &\equiv & \frac{1}{4\pi} \left[ \frac{\omega '}{r^2}e^{-\Lambda}
+ \frac{(1 - \omega ^2)^2}{r^4} + \frac{\phi ^2}{4r^2} (1 + \omega ^2)
\right].
\label{fourteena}
\eea
Substituting (\ref{fourteena}) in (\ref{thirteena}) yields
\be
   T_r^r (r) = \frac{e^{-\frac{\Gamma}{2}}}{r^2}\int _{r_h}^r
\left\{ -
(e^{\frac{\Gamma}{2}}r^2)' {\cal E} + \frac{2}{r} {\cal J}
\right\} dr
\label{fifteena}
\ee
\be
     (T_r^r)' (r) = -\frac{e^{-\frac{\Gamma}{2}}}{r^2}
 (e^{\frac{\Gamma}{2}}r^2)'
{\cal F} + \frac{2}{r}{\cal J}
\label{fifteenb}
\ee
where ${\cal E}$ is expressed as
\be
   {\cal E} =\frac{1}{4\pi} \left[  \frac{(\omega ')^2}{r^2}
e^{-\Lambda } + \frac{(1-\omega ^2)^2}{2r^4} +
\frac{\phi ^2(1 + \omega )^2}{4r^2} + \frac{1}{2} (\phi ')^2e^{-\Lambda }
+ \frac{{\lambda}}{4} (\phi ^2 - { v}^2)^2  \right].
\label{sixteena}
\ee
We now turn to the Einstein equations for the first time, following
the analysis of ref. \cite{bekmod}. Our aim is to examine whether there is a
contradiction with the requirement
of the non-negative energy density. These equations read for our system
\bea
   e^{-\Lambda} (r^{-2} - r^{-1}\Lambda ' ) - r^{-2} &=&
8\pi T_t^t = -8\pi {\cal E} \nn \\
e^{-\Lambda} (r^{-1} \Gamma' + r^{-2} ) - r^{-2} &=& 8\pi T_r^r.
\label{seventeena}
\eea
Integrating out the first of these yields
\be
          e^{-\Lambda }  = 1 -\frac{8\pi}{r} \int _{r_h}^r
{\cal E} r^2 dr - \frac{2 {\cal M_{0}}}{r}
\label{eighteena}
\ee
where ${\cal M_{0}}$ is a constant of integration.
\pr
The requirement for
asymptotic flatness of space-time implies the following
asymptotic behaviour for the energy-density functional ${\cal E} \sim
O(r^{-3}) $ as $r \rightarrow \infty $, so that $\Lambda \sim O(r^{-1}) $.
In order that $e^{\Lambda} \rightarrow \infty $ at the horizon,
$r \rightarrow r_h $, ${\cal M_{0}}$ is fixed by
 \be
               {\cal M_{0}} = \frac{r_h}{2}.
\label{nineteen}
\ee
The second of the equations (\ref{seventeena}) can be rewritten in the form
\be
 e^{-\frac{\Gamma}{2}} r^{-2}
 (r^2 e^{\frac{\Gamma}{2}})' = \left[ 4\pi r T_r^r
+ \frac{1}{2r} \right] e^{\Lambda} + \frac{3}{2r}.
\label{twenty}
\ee
\pr
Consider, first, the behaviour of $T_r^r$ as $ r \rightarrow \infty $.
Asymptotically, $e^{\frac{\Gamma}{2}} \rightarrow 1 $, and so the leading
behaviour of $(T_r^r)'$ is
\be
                   (T_r^r) ' = \frac{2}{r} [{\cal J} - {\cal F} ].
\label{21}
\ee
We, now, note that the fields $\omega $ and $\phi $ have masses
$\frac{{ v}}{2}$ and $\mu  = \sqrt{{ \lambda}}{ v}$
respectively. From the field equations and the requirement of finite energy
density their behaviour at infinity must then be
\bea
      \omega (r) &\sim & -1 + ce^{-\frac{{ v}}{2} r}  \nn \\
      \phi (r)  &\sim & { v} + a e^{-\sqrt{2} \mu r}
\label{22}
\eea
for some constants $c$ and $a$. Hence, the leading asymptotic behaviour
of ${\cal J}$ and ${\cal F}$ is
\bea
  {\cal J} &\sim & \frac{1}{4\pi} \left[ \frac{c^2 { v}^2}{4r^2}e^{-
{ v}r} + \frac{2c^2}{r^4} e^{-{ v}r} +
\frac{{ v}^2 c^2 }{4r^2} e^{-{ v}r} \right] \nn \\
{\cal F} &\sim & \frac{1}{4\pi} \left[ \frac{c^2 { v}^2}
{2r^2} e^{-{ v} r} + 2 a^2 \mu ^2 e^{-\sqrt{2} \mu r} \right]
\label{23}
\eea
since $e^{-\Lambda } \rightarrow 1$ asymptotically.
\pr
The leading behaviour of $(T_r^r)'$, therefore, is
\be
   (T_r^r)' \sim \frac{1}{4\pi} \left[ \frac{2c^2}{r^4} e^{-{ v}r}
- 2a^2 \mu ^2 e^{-2\sqrt{2} \mu r} \right].
\label{24}
\ee
There are two possible cases: (i) $2\sqrt{2}\mu > { v} $ (corresponding
to ${ \lambda } > 1/8$); in this case $(T_r^r)' > 0$
asymptotically, (ii) $2\sqrt{2} \mu \le { v}$ (corresponding to
${ \lambda } \le 1/8$) ; then, $(T_r^r)' < 0$ asymptotically.
\pr
Since ${\cal J}$ vanishes exponentially at infinity, and ${\cal E} \sim
O[r^{-3}]$ as $r \rightarrow \infty$, the integral
defining $T_r^r (r)$ converges as $r \rightarrow \infty $
and $|T_r^r|$ decreases as $r^{-2}$.
\pr
Thus, in case (i) above, $T_r^r$ is negative and increasing as $r \rightarrow
\infty$, and in case (ii) $T_r^r$ is positive and decreasing.
\pr
Now turn to the behaviour of $T_r^r$ at the horizon.
When $r \simeq r_h$ , ${\cal E}$ and ${\cal J}$ are both finite, and
$\Gamma'$ diverges as $1/(r- r_h)$.
Thus the main contribution to $T_r^r$ as $r \simeq r_h$ is
\be
   T_r^r (t) \simeq \frac{e^{-\Gamma/2}}{r^2} \int _{r_h} ^r
(-e^{\Gamma/2}
r^2) \frac{\Gamma'}{2} {\cal E} dr
\label{25}
\ee
which is finite.
\pr
At the horizon, $e^\Gamma = 0$; outside the horizon, $e^\Gamma > 0$ .
Hence $\Gamma' >0$ sufficiently close to the horizon,
 and, since ${\cal E} \ge 0$,
$T_r^r < 0$ for $r$ sufficiently close to the horizon.
\pr
Since ${\cal F} \sim O[r-r_h]$ at $r \simeq r_h$, $(T_r^r)'$ is finite
at the horizon and the leading contribution is
\be
  (T_r^r)' (r_h) \simeq -\frac{\Gamma'}{2} {\cal F} + \frac{2}{r}{\cal J}.
\label{26}
\ee
{}From ref. \cite{greene} we record the relation
\be
re^{-\Lambda} \frac{\Gamma'}{2}   =  e^{-\Lambda}\left[ \omega '^{2} +
\frac{1}{2} r^2 (\phi ')^2 \right] - \frac{1}{2} \frac{(1-\omega ^2)^2}{r^2}
 - \frac{1}{4}\phi ^2 (1 + \omega ^2)^2 + \frac{m}{r} -
\frac{{ \lambda}}{4} (\phi ^2 - { v}^2)^2 r^2
\label{27}
\ee
where $e^{-\Lambda} = 1 - \frac{2m(r)}{r}$. Hence,
\bea
(T_r^r)' &=& -\frac{e^{-\Lambda}}{4\pi r}\left[ \frac{2(\omega ')^2}{r^2}
+ (\phi ')^2 \right] \left\{ (\omega ')^2 + \frac{1}{2} r^2 (\phi ')^2
- \frac{1}{2} e^\Lambda \frac{(1-\omega ^2)}{r^2}  \right.\\
 & & \left.
-\frac{\phi ^2}{4} e^\Lambda (1 + \omega )^2 + \nn
\frac{m}{r}e^\Lambda  \frac{{\lambda}}{4} (\phi ^2 -
{ v}^2 )r^2 e^\Lambda \right\} \\
 & &
+ \frac{1}{2\pi r} \left[ \frac{(\omega ')^2}{r^2} e^\Lambda +
\frac{(1 - \omega ^2)^2}{r^4}  + \frac{\phi ^2}{4r^2}
(1 + \omega)^2 \right].
\label{28}
\eea
For $r \simeq r_h$,
this expression simplifies to
\bea
(T_r^r)'(r_h) & \simeq & {\cal J}(r_h) \left[ \frac{2}{r_h} + \frac{4\pi}{r_h}
{\tilde {\cal F}}(r_h) \right] \nn  \\
 & &  - {\tilde {\cal F}}(r_h) \left[ \frac{1}{2}
+ \frac{1}{2}\frac{(1 - \omega ^2)^2}{r_h^2} -
\frac{{ \lambda}}{4}(\phi ^2 - { v}^2)^2 r_h^2 \right] \nn  \\
 & = &
{\tilde {\cal F}}(r_{h}) \left[
\frac {(1- \omega ^{2})^{2}}{2r_{h}^{3}} +
\frac {\phi ^{2}}{4r_{h}}(1+\omega ^{2})^{2} +
\frac {\lambda }{4}r_{h}(\phi ^{2} -v^{2})^{2} -
\frac {1}{2r_{h}} \right] \nn \\ & &
+ \frac {2}{r_{h}} {\cal J}
\label{29}
\eea
where ${\tilde {\cal F}} = e^\Lambda {\cal F}(r) = \frac{1}{4\pi}
[\frac{2(\omega ')^2}{r^2} + (\phi')^2]$.
\pr
Consider for simplicity the case $r_{h}=1$.
Then, from the field equations \cite{greene} (see  also Appendix A)
\bea
\omega _{h}' & = &
\frac {1}{\cal D} \left[
\frac {1}{4} \phi _{h}^{2} (1+\omega _{h}) -
\omega _{h} (1-\omega _{h}^{2}) \right] =
\frac {\cal A}{\cal D} \\
\phi _{h}' & = &
\frac {1}{\cal D} \left[
\frac {1}{2} \phi _{h} (1+\omega _{h})^{2}
+\lambda \phi _{h} (\phi _{h}^{2}- v^{2}) \right] =
\frac {\cal B}{\cal D}
\eea
where
\be
{\cal D}=1- (1-\omega _{h}^{2})^{2} -
\frac {1}{2} \phi _{h}^{2} (1+\omega _{h})^{2}
- \frac {1}{2} \lambda (\phi _{h}^{2} -v^{2})^{2}.
\ee
Then the expression (\ref{29}) becomes
\be
(T^{r}_{r})'(r_{h})=
\frac {1}{4\pi {\cal {D}}} \left[
8\pi {\cal D} {\cal J} - {\cal A}^{2} -\frac {1}{2} {\cal B}^{2}
\right]
=\frac {{\cal {C}}}{ 4  \pi {\cal {D}}}.
\ee
{}From the field equations \cite{greene} (see Appendix A),
\be
{\cal D}=1-2m_{h}'
\ee
which is always positive because the black holes are non-extremal.
(See Appendix A for further discussion of this point.)
Thus the sign of $(T^{r}_{r})'(r_{h})$ is the same as that of
${\cal C}$.
Simplifying, we have
\be
{\cal C}=c_{1}+c_{2}+c_{3}+c_{4}+c_{5}
\ee
where
\bea
c_{1} & = &
(1-\omega _{h}^{2})^{2} \omega _{h}^{2} (3-2\omega _{h}^{2}) \\
c_{2} & = &
\frac {1}{8} \phi _{h}^{2} (1+\omega _{h}) (1-\omega _{h}^{2})
(12\omega _{h}^{3} +12\omega _{h}^{2} -7\omega _{h}-9) \\
c_{3} & = &\frac {1}{16} \phi _{h}^{4} (1+\omega _{h})^{2}
(4\omega _{h}^{2} +8\omega _{h}+5) \\
c_{4} & = &
-\lambda (\phi _{h}^{2}-v^{2})^{2} \left[
(1-\omega _{h}^{2})^{2} +\frac {1}{2} \lambda \phi _{h}^{2} \right] \\
c_{5} & = &
\frac {1}{4} \lambda \phi _{h}^{2} (v^{2}-\phi _{h}^{2})
(1+\omega _{h})^{2} (2-v^{2}+\phi _{h}^{2}).
\eea
The first term is always positive since $|\omega _{h}|\le 1$.
The cubic in $c_{2}$ possesses a local maximum at $\omega _{h}=-0.886$
where it has the value $-1.724$ and a local minimum at
$\omega _{h}=0.219$ where it equals $-9.831$.
The cubic has a single root at $\omega _{h}=0.8215$, and thus is
positive for $\omega _{h}>0.8215 $ and negative for
$\omega _{h}<0.8215$.
The quadratic in $c_{3}$ is always positive and possesses no real
roots.
It has a minimum value of 1 when $\omega _{h}=-1$.
The term $c_{4}$ is always negative and $c_{5}$ is always positive
since $|\phi _{h}|\le v$.
\pr
In order to assess whether or not $\cal C$ as a whole is positive or
negative, we shall consider each branch of black hole solutions in
turn.
\pr
Firstly, consider the $k=1$ branch of solutions.
As $v$ increases from 0 up to $v_{max}=0.352$, $\omega _{h}$
increases monotonically from 0.632 to 0.869.
The derivative of the first term is
\be
\frac {dc_{1}}{d\omega _{h}}=
2\omega _{h}(1-\omega _{h}^{2})(3-13\omega _{h}+8\omega _{h}^{2})
\ee
where the quartic has roots given by $\omega _{h}^{2}=1.347 {\mbox { or }}
0.278$.
This derivative is negative for $\omega _{h}\in (0.632,0.869) $
and hence the first term decreases as $v$ increases,  and is bounded below
by its value at the bifurcation point $v=v_{max}$, namely
\be
c_{1}\ge 0.0674.
\ee
The cubic in $c_{2}$ increases as $\omega _{h}$ increases from 0.632
to 0.869, and is bounded below by its value when $\omega _{h}=0.632$
\be
12\omega _{h}^{3} +12\omega _{h}^{2} -7 \omega _{h} -9
\ge -5.602.
\ee
 Along this branch of solutions, it is also true that
\bea
(1-\omega _{h}^{2}) & \le & 1-(0.632)^{2}=0.601 \nn  \\
1+\omega _{h} & \le & 2 \nn  \\
\phi _{h} & \le & 0.19v \le 0.0669.
\eea
Altogether this gives
\be
c_{2}\ge -5.602 \times \frac {1}{8} \times (0.0669)^{2}
\times 2\times 0.601 =
-3.767 \times 10^{-3}.
\ee
The quadratic in $c_{3}$ increases as $\omega _{h}$ increases
along the branch and so is bounded above by its value when
$\omega _{h}=0.869$;
\be
4\omega _{h}^{2} +8\omega _{h} +5 \le 14.973.
\ee
Thus,
\be
c_{3}\ge -\frac {1}{16} \times (0.0669 )^{4} \times 2^{2}
\times 14.973 =
-7.498 \times 10^{-5}.
\ee
For the fourth term, since
\be
(\phi ^{2}_{h} -v^{2})^{2}\le v^{4}
\ee
we have
\be
c_{4} \ge -0.15 \times (0.352)^{4} \times \left(
(0.601)^{2} +\frac {1}{2} \times 0.15\times (0.0669)^{2}
\right) =-8.326 \times 10^{-4}
\ee
and finally,
\be
c_{5}\ge 0.
\ee
Thus, adding these expressions up, one obtains
\be
{\cal C}\ge 0.0627 \ge 0
\ee
so that $(T^{r}_{r})'(r_{h})>0$ along the whole of the
$k=1$ branch of black hole solutions.
\pr
For the quasi-$k=0$ branch, as $v$ decreases from $v_{max}$ down
to 0, $\omega _{h}$ increases monotonically from 0.869, subject to
the inequality
\be
0.869<\omega _{h} < 1-0.1v^{2}.
\ee
Hence, along this branch,
\be
(1-\omega _{h}^{2})^{2} =(1-\omega _{h})^{2} (1+\omega _{h})^{2}
\ge (0.1v^{2})^{2} \times (1.869)^{2}
=0.0349 v^{4}.
\ee
Thus the first term is bounded below as follows:
\be
c_{1} \ge 0.0349 v^{4} \times (0.869)^{2} \times 1
= 0.0264v^{4}.
\ee
Since $\omega _{h}>0.869$, the cubic in $c_{2}$ is positive all
along this branch, so that $c_{2}$ and $c_{5}$ are positive.
The quadratic in $c_{3}$ is bounded above by its value when
$\omega _{h}=2$, i.e.
\be
4\omega _{h}^{2} +8\omega _{h} +5 \le 37.
\ee
All along the quasi-$k=0$ branch,
\be
\phi _{h} \le 0.5 v^{2}.
\ee
This gives
\be
c_{3}  \ge -\frac {1}{16} \times (0.5v^{2})^{4} \times 2^{2}
\times 37 = -0.578 v^{8}.
\ee
Since $v\le 0.352$,
\be
c_{3} \ge -0.578 \times (0.352)^{4} \times v^{4}
=-8.874\times 10^{-3} v^{4}.
\ee
Finally, for $c_{4}$ we have
\bea
c_{4} & \ge &
-0.15 v^{4} \left[ \frac {1}{2} \times 0.15 \times (0.5)^{2} v^{4}
+(1-(0.869)^{2})^{2} \right] \nn  \\
 & = &
-8.992 \times 10^{-3}v^{4} -2.813 \times 10^{-3} v^{8} \nn  \\
 & \ge &
-8.992 \times 10^{-3} v^{4} -2.813 \times 10^{-3} \times
(0.352)^{4} v^{4} \nn  \\
 & = &
-9.035 \times 10^{-3} v^{4}.
\eea
In total this gives
\be
{\cal C} \ge 0.0264v^{4} -8.874\times 10^{-3} v^{4}
-9.035 \times 10^{-3} v^{4} = 8.491 \times 10^{-3}v^{4}
\ge 0.
\ee
In conclusion, $(T^{r}_{r})'(r_{h})$ is positive for all the black
hole solutions having one node in $\omega $, regardless of the
value of the Higgs mass $v$.
\pr
Let us now check on possible contradictions
with Einstein's equations.
\pr
Consider first the case
${ \lambda} > 1/8$. Then, as $r \rightarrow
\infty$, $T_r^r < 0 $ and $(T_r^r)' > 0$.
As $r \rightarrow r_h $, $T_r^r < 0$ and $(T_r^r)' > 0$.
Hence there is no contradiction with
Einstein's equations in this case.
\pr
Consider now the case $\lambda \le 1/8$.
In this case, as $r \rightarrow \infty$, $T_r^r > 0$ and
$(T_r^r)' < 0$, whilst as $r \rightarrow r_h$, $T_r^r
< 0 $ and $(T_r^r)' > 0$.
Hence, there is an interval $[ r_a, r_b ]$ in which $(T_r^r)' $ is positive
and there exists a `critical' distance $r_c \in (r_a, r_b)$ at which
$T_r^r$ changes sign.
\pr
However, unlike the case when the gauge fields are absent~\cite{bekmod},
here
there is {\it no contradiction} with the result following from Einstein
equations, because $(T_r^r)' > 0$ in some open interval close to the
horizon, as we have seen above.
\pr
In conclusion the method of ref. \cite{bekmod} cannot be used to prove
a `no-scalar-hair' theorem for the EYMH system, as expected from the
existence of the
explicit solution of ref. \cite{greene}. The {\it key} difference is
the presence of the positive term $\frac{2}{r}{\cal J}$
in the expression (\ref{fifteenb}) for $(T_r^r)'$. This term is
dependent on the Yang-Mills field and vanishes if this field is absent.
Thus, there is a sort of `balancing' between the gravitational
attraction and the non-Abelian gauge field repulsion, which
is responsible for the existence of the classical non-trivial
black-hole solution of ref. \cite{greene}.
However, as we shall discuss below, this solution is not stable against
(linear) perturbations of the various field configurations~\cite{winst}.
Thus, although the `letter' of the `no-scalar-hair' theorem of ref.
\cite{bekmod}, based on non-negative scalar-field-energy density,
is violated, its `spirit' is maintained in the sense that there exist
instabilities which that the solution cannot be formed
as a result of collapse of stable matter.

\section{Instability analysis of
sphaleron sector of the EYMH black hole}
\pr
The black hole solutions of ref. \cite{greene} in the EYMH system
resemble the sphaleron solutions
in $SU(2)$ gauge theory and one would expect them to be
unstable for topological reasons. Below we shall confirm this
expectation by proving~\cite{winst} the existence of unstable
modes in the sphaleron sector
of the EYMH black hole system (for notation and definitions
see Appendix A).
\pr
Recently, an instability
proof of sphaleron solutions for arbitrary gauge
groups in the EYM system has been given \cite{bs,brodbeck}.
The method consists of studying linearised radial perturbations
around an equilibrium solution, whose detailed knowledge
is not necessary to establish stability.
The stability is examined by mapping the system
of algebraic equations for the perturbations
into a coupled system of differential equations
of Schr\"odinger type \cite{bs,brodbeck}.
As in the particle case of ref. \cite{bartnik}, the
instability of the solution is established once
a bound state in the respective
Schr\"odinger equations is found.
The latter shows up as an
imaginary frequency mode in the spectrum, leading to an
exponentially growing mode.
There is an elegant physical interpretation behind this
analysis, which is similar to the
Cooper pair instability of super-conductivity.
The gravitational attraction balances the
non-Abelian gauge field repulsion in the classical
solution \cite{bartnik}, but the existence of bound states
implies imaginary parts in the quantum ground state
which lead to instabilities of the solution, in much
the same way as the classical ground state
in super-conductivity is not the absolute minimum
of the free energy.
\pr
However, this method cannot be applied directly to the black
hole case, due to divergences occuring in some of the
expressions involved. This is
a result of the singular behaviour
of the metric function at the physical space-time boundaries
(horizon) of the black hole.

\subsection{Linearized perturbations and instabilities}
\pr
It is the purpose of this section to generalise the method
of ref. \cite{bs} to incorporate the black hole solution
of the EYMH system of ref. \cite{greene}. By
constructing appropriate
trial linear radial perturbations,
following ref. \cite{brodbeck,volkov},
we show the existence of bound states in the
spectrum of the coupled Schr\"odinger
equations, and thus the instability of the black hole.
Detailed knowledge of the black hole solutions is not
actually required, apart from the fact that
the existence of an horizon leads to modifications
of the trial perturbations as compared to those of
ref. \cite{bs,brodbeck},  in order to avoid divergences
in the respective expressions \cite{volkov}.
\pr
We start by sketching the basic steps \cite{bs,volkov}
that will lead to a study of the stability of
a classical solution $\phi _s (x, t)$
with finite energy
in a (generic) classical field theory.
One considers
small perturbations $\delta \phi (x,t)$
around $\phi _s  (x, t)$, and specifies \cite{bs}
the time-dependence as
\be
    \delta \phi (x ,t ) = \exp (-i \Omega t ) \Psi (x ).
\label{linear}
\ee
The
linearised
system (with respect to such perturbations),
obtained from the equations of motion,
can be
cast into a Schr\"odinger eigenvalue problem
\be
   {\cal H} \Psi = \Omega ^2 A \Psi
\label{schr}
\ee
where the operators ${\cal H}$, $A$ are assumed independent
of the `frequency' $\Omega$. As we shall show later on, this is
indeed the case of our black hole solution of the EYMH system.
In that case it will also be shown that
${\cal H}$ is a self-adjoint operator with respect
to a properly defined inner (scalar) product in the space
of functions $\{\Psi \}$ \cite{bs},
and the $A$ matrix is positive definite,
$ <\Psi | A | \Psi > > 0 $.
A criterion for instability is the existence of an imaginary
frequency  mode in (\ref{schr})
\be
     \Omega ^2 < 0.
\label{inst}
\ee
This is usually difficult to solve analytically
in realistic models,
and usually numerical calculations are required \cite{stab}.
A less informative method which admits analytic treatment
has been proposed recently in ref. \cite{bs,volkov},
and we shall follow this
 for the purposes of the present work.
The method consists of a variational approach
which makes use of the following functional
defined through (\ref{schr}):
\be
   \Omega ^2 (\Psi ) = \frac{ <\Psi | {\cal H } | \Psi >}
{<\Psi | A | \Psi >}
\label{funct}
\ee
with $\Psi $ a {\it trial} function.
The lowest eigenvalue is known to provide a
{\it lower} bound for this functional.
Thus,
the criterion of instability, which is equivalent to
(\ref{inst}), in this approach
reads
\bea
  \Omega ^2 ( \Psi ) &<& 0
\nn \\
    <\Psi | A | \Psi > &<& \infty .
\label{trueinst}
\eea
The first of the above
conditions implies that the operator
${\cal H }$ is not positive definite, and therefore
negative eigenvalues do exist.
The second condition, on the {\it finiteness}
of the expectation value of the operator $A$,
is required to ensure that $\Psi$ lies in the
Hilbert space containing the domain of ${\cal H}$.
 In certain cases, especially
in the black hole case,
there are divergences
due to singular behaviour of modes at, say, the horizons,
which could spoil these conditions
(\ref{trueinst}).
The advantage of the above variational method
lies in the fact that it is an easier task
to choose appropriate trial functions $\Psi $
that satisfy (\ref{trueinst}) than solving the
original eigenvalue problem (\ref{schr}).
In what follows we shall apply this second method
to the black hole solution of ref. \cite{greene}.
\pr
For completeness,
we first review basic formulas
of the spherically symmetric black hole solutions
of the EYMH system \cite{greene}.
The space-time metric
takes the form \cite{greene}
\be
ds^2 =-N(t,r) S^{2}(t,r) dt^2 + N^{-1} dr^2 +
r^2 (d\theta ^2 + \sin^2 \theta d\varphi ^2)
\label{one}
\ee
and we assume the following ansatz for
the non-Abelian gauge potential \cite{greene,bs}
\be
A  = a_0 \tau _r dt + a_1 \tau _r dr + (\omega +1 )
[- \tau _\varphi d\theta + \tau _\theta \sin \theta d\varphi ]
+ {\tilde \omega} [ \tau _\theta d\theta + \tau _\varphi \sin \theta d\varphi ]
\label{two}
\ee
where $\omega, {\tilde \omega}$ and $a_i, i = 0,1 $
are functions of $t, r$. The $\tau _i$ are appropriately normalised
spherical
generators of the SU(2) group in the notation
of ref. \cite{bs}.
\pr
The Higgs doublet assumes the form
\be
{\tilde \Phi} \equiv \frac{1}{\sqrt{2}} \left( \begin{array}{c}
\nonumber  \psi _2  + i \psi _1 \\
           \phi - i \psi _3  \end{array}\right)
\qquad ; \qquad {\mbox {\boldmath $ \psi $}}
 = \psi {\mbox {\boldmath ${\hat r}$}}
\label{three}
\ee
with the Higgs potential
\be
   V({\tilde \Phi } )=\frac{\lambda }{4} ( {\tilde \Phi} ^{\dag}
{\tilde \Phi }  - { v}^2)^2
\label{four}
\ee
where ${ v}$ denotes the v.e.v. of ${\tilde \Phi } $ in the non-trivial
vacuum.
\pr
The quantities $\omega, \phi$ satisfy the static field
equations
\bea
  N \omega '' + \frac{(N S)'}{S} \omega ' & =& \frac{1}{r^2}
(\omega ^2 - 1) \omega + \frac{\phi ^2}{4}
(\omega + 1)    \nn \\
N \phi '' + \frac{(N S)'}{ S} \phi ' + \frac{2N}{r} \phi '
&=& \frac{1}{2r^2} \phi (\omega +1 )^2 + \lambda \phi
(\phi ^2 - {v}^2 )
\label{five}
\eea
where the  prime denotes differentiation with respect
to $r$. For later use, we also mention that
a dot
will denote
differentiation
with respect to $t$.
\pr
If we choose a gauge in which $\delta a_{0} =0$,
the linearised perturbation equations decouple into two sectors
\cite{bs} . The first consists of the gravitational modes
$\delta N$, $\delta S$, $\delta \omega$ and
$\delta \phi$ and the second of the matter perturbations
$\delta a_{1}$, $\delta {\tilde {\omega }}$
and $\delta \psi $.
For our analysis in this section it will be sufficient
to concentrate on the matter perturbations, setting the
gravitational perturbations  $\delta N$ and $\delta S$
to zero, because an instability will show up in
this sector of the theory. An instability study in the gravitational
sector will be discussed in the following section 4.
The equations for
the linearised matter perturbations
take the form \cite{bs}
\be
   {\cal H} \Psi + A {\ddot \Psi } = 0
\label{six}
\ee
with,
\be
\Psi = \left( \begin{array}{c}
\nonumber  \delta a_1  \\
\nonumber  \delta {\tilde \omega}  \\
\nonumber  \delta \psi    \end{array}\right)
\label{seven}
\ee
and,
\be
A = \left( \begin{array}{ccc}
  Nr^2 &   0  & 0 \\
  0 &  2   & 0 \\
  0 &  0   & r^2  \\
\end{array}\right)
\label{eight}
\ee
and the components of ${\cal H}$ are
\bea
{\cal H}_{a_1a_1} &=& 2 (N S)^2 \left( \omega ^2
+ \frac{r^2}{8} \phi ^2 \right) \nn \\
{\cal H}_{{\tilde \omega} {\tilde \omega}} &=&
2 p_* ^2 + 2NS^2 \left( \frac{\omega ^2 -1}{r^2} + \frac{\phi ^2}{4}
\right)
\nn \\
{\cal H}_{\psi\psi} &=& 2 p_*\frac{r^2}{2} p_* +
2 NS^2 \left( \frac{(-\omega + 1)^2}{4} + \frac{r^2 }{2}\lambda
(\phi ^2 - {v}^2) \right) \nn \\
{\cal H}_{a_1{\tilde \omega}} &=& - 2i N S [ (p_* \omega) - \omega p_* ]
\nn \\
{\cal H}_{{\tilde \omega} a_1} &=&
 -2i [ p_* N S \omega + N S (p_* \omega) ]
\\
{\cal H}_{a_1 \psi } &=& \frac{i r^2}{2} N S [(p_* \phi) - \phi p_* ]
\nn \\
{\cal H}_{\psi a_1} &=& i p_* \frac{r^2}{2}
NS \phi + i\frac{r^2}{2} NS (p_* \phi )
\nn \\
{\cal H}_{{\tilde \omega}\psi} &=& {\cal H}_{\psi {\tilde
\omega}} = -\phi N S^2     \nn
\label{nine}
\eea
where the operator $p_*$ is
\be
    p_* \equiv - i NS \frac{d}{dr}.
\label{ten}
\ee
Upon specifying the time-dependence
(\ref{linear})
\be
  \Psi (r, t) = \Psi (r) e^{i \sigma t}  \qquad ;
\qquad
\Psi (r) = \left( \begin{array}{c}
\nonumber  \delta a_1 (r) \\
\nonumber  \delta {\tilde \omega} (r) \\
\nonumber  \delta \psi (r)   \end{array}\right)
\label{linar2}
\ee
one arrives easily to an eigenvalue
problem of the form (\ref{schr}), which can then be
extended to the variational approach (\ref{trueinst}).
\pr
To this end, we choose
as trial perturbations the following expressions
(c.f. \cite{bs})
\bea
\nonumber  \delta a_1 &=&- \omega ' Z  \\
\nonumber  \delta {\tilde \omega} &=& (\omega ^2 - 1) Z  \\
\nonumber  \delta \psi &=& -\frac{1}{2} \phi (\omega + 1) Z
\label{eleven}
\eea
where $Z$ is a function of $r$ to be determined.
\pr
One may define the inner product
\be
  <\Psi | X > \equiv \int _{r_h} ^{\infty} {\overline \Psi } X
\frac{1}{NS} dr
\label{twelve}
\ee
where $r_h$ is the position of the horizon of the black hole.
The operator ${\cal H}$ is then symmetric with respect to
this scalar product.
Following ref. \cite{bs}, consider the expectation value
\be
  <\Psi | A | \Psi > = \int _{r_h}^{\infty}
dr \frac{1}{NS} Z^2 \left[ Nr^2 (\omega ')^2 +
2 (\omega ^2 - 1)^2 +
\frac{r^2}{4} \phi ^2 (\omega + 1)^2 \right]
\label{thirteen}
\ee
which is clearly positive definite for real $Z$.
Its finiteness will be examined later, and depends
on the choice of the function $Z$.
\pr
Next, we proceed to the evaluation of
the expectation value of the
Hamiltonian ${\cal H}$ (\ref{nine}); after a tedious calculation
one obtains
\bea
  <\Psi | {\cal H} | \Psi > &=& \int _{r_h}^{\infty}
dr S Z^2 \{ - 2 N (\omega ')^2 + 2 P^2 N (\omega ^2 - 1)^2
 \nn \\  &  &
+ \frac{1}{4} P^2 N r^2 \phi ^2 (\omega + 1)^2
- \frac{2}{r^2} (\omega ^2 - 1)^2  - \frac{1}{2}
\phi ^2 (\omega + 1)^2 \}  \\
\nn  & + & {\mbox {boundary terms}}
\label{fourteen}
\eea
where $P \equiv \frac{1}{Z} \frac{d Z}{d r} $.
The boundary terms will be shown to vanish so we omit them in
the expression (\ref{fourteen}). The final result is
\bea
<\Psi | {\cal H} | \Psi >  &= & \int _{r_h}^{\infty}
dr S \left\{ - 2 N (\omega ')^2 - \frac{2}{r^2}
(\omega ^2 - 1)^2 - \frac{1}{2} \phi ^2 ( \omega + 1)^2   \right\}
 \nn \\
 & + & \int _{r_h}^{\infty} dr \left\{
\frac{2}{r^2} (\omega ^2 - 1) ^2 + \phi ^2 (\omega + 1)^2
+ 2 N (\omega ')^2 \right\} S ( 1 - Z^2)  \nn \\
 & +  & \int _{r_h}^{\infty} dr S N \left( \frac{d Z}{dr} \right) ^2
\left[ 2 (\omega ^2 -1)^2 + \frac{1}{4} r^2 \phi ^2 (\omega + 1)^2
\right] .
\label{fifteen}
\eea
\pr
The first of these terms is manifestly negative.
To examine the remaining two, we introduce the
`tortoise' co-ordinate $r^*$ defined by \cite{volkov}
\be
      \frac{d r^*}{dr} = \frac{1}{N S}
\label{tortoise}
\ee
 and define a sequence of functions $Z _k ( r^* )$ by
\cite{volkov}
\be
  Z_k ( r^* ) = Z\left( \frac{r^*}{k}
\right) \qquad ; \qquad k =1,2, \ldots
\label{seventeen}
\ee
where
\bea
Z ( r^* ) &=& Z ( -r^* ), \nn \\
Z ( r^* ) &=& 1 \qquad
{\mbox {for $ r^* \in [ 0, a] $}}  \nn \\
 - D \le&  \frac{d Z}{d r^*} & < 0, \qquad
{\mbox { for $r^* \in [a, a + 1 ]$}}
\nn \\
 Z( r^* ) &=& 0  \qquad
{\mbox {for $ r^* > a + 1 $}}
\label{eighteen}
\eea
where $a$, $D$ are arbitrary positive constants.
Then, for each value of $k$ the vacuum expectation values
of ${\cal H }$ and $A$ are finite,
$ <\Psi | {\cal H } | \Psi > < \infty $,
and $<\Psi |A| \Psi > < \infty $,
with $Z = Z_k $, and all boundary terms vanish. This
justifies {\it a posteriori} their being dropped in
eq. (\ref{fourteen}).  The integrands in the second and third terms
of eq. (\ref{fifteen}) are uniformly convergent
and tend to zero as $ k \rightarrow \infty $. Hence, choosing
$k $ sufficiently large the dominant contribution
in (\ref{fifteen}) comes from the first term which is negative.
\pr
This confirms the existence of bound states in the
Schr\"odinger equation (\ref{six}), (\ref{schr}),
and thereby the
instability (\ref{trueinst})
of the associated black hole solution of ref. \cite{greene}
in the coupled EYMH system.
The above analysis reveals the existence of at least one
negative {\it odd-parity} eigenmode
in the spectrum
of the
EYMH black hole.
\pr
\subsection{Counting sphaleron-like unstable modes in the EYMH
system}
\pr
The exact number of such negative modes is an interesting
question and we next proceed to investigate it.
Recently, a method for determining the number of the
sphaleron-like unstable modes
has been applied
by Volkov et al. \cite{eigenmodes}
to the gravitating sphaleron case. We have been able
to extend it to the present EYMH black hole.
The method consists of mapping the system
of linearized perturbations to a system of coupled
Scr\"odinger-like equations. Counting of unstable
modes is then equivalent to counting bound states
of the quantum-mechanical analogue system.
It is important to notice that
due to the fact that the EYMH black hole solution
is not known analytically, but only
numerically, it will be necessary to make
certain physically plausible assumptions
concerning certain
analyticity requirements~\cite{simon} for the
solutions of the analogue system. This is equivalent
to requiring that
the conditions for the
validity of perturbation theory
in ordinary quantum mechanics
be applied to this problem.
Details are described below.
\pr
Working in the gauge $\delta a_0 = 0$, and denoting the
derivative with respect to the tortoise coordinate (\ref{tortoise})
by a prime,
we can write
the linearised perturbations in the sphaleron sector of the EYMH system
as~\cite{bs}:
\bea
 2N^2 S^2 \left( \omega ^2 + \frac{r^2}{8}\phi ^2 \right) \delta a_1
+ 2 NS (\omega
\delta {\tilde \omega }' - \omega ' \delta {\tilde \omega } ) & & \nn \\
+ \frac{1}{2} r^2NS (\phi ' \delta \psi - \phi \delta \psi ' ) &=&
Nr^2 \sigma ^2 \delta a_1 \nn \\
2(N S \omega \delta a_1 )' + 2 N S \omega ' \delta a_1 +
\delta {\tilde \omega } '' + N S^2 \phi \delta \psi  & & \nn \\
-\frac{2}{r^2} S^2 \left( (\omega ^2 - 1)
+ \frac{\phi ^2}{4}\right) \delta {\tilde \omega}
&=& -2\sigma ^2 \delta {\tilde \omega } \nn \\
\frac{1}{2} (NSr^2 \phi \delta a_1 )' + \frac{1}{2}r^2 N S \phi ' \delta a_1
- N S^2 \phi \delta {\tilde \omega } - (r^2 \delta \psi ')'  & &  \nn \\
+ 2 N S^2 \left( \frac{(1-\omega)^2}{4} + \frac{1}{2}r^2 \lambda
(\phi ^2 - { v}^2 ) \right) \delta \psi &=& r^2 \sigma^2 \delta \psi
\label{3-1}
\eea
together with the Gauss constraint
\be
 \sigma ^2 \left\{  \left(  \frac{r^2}{S} \delta a_1 \right)'
+ 2 \omega {\tilde \omega}
-r^2 \frac{\phi}{2} \delta \psi \right\} = 0.
\label{3-2}
\ee
Define  $\delta \xi = r \delta \psi $, $\delta \alpha = \frac{r^2}{2S}
\delta a_1 $. Then
\be
   \sigma ^2 \delta \alpha = f(r^*)
\label{3-3}
\ee
where
\be
f(r^*) = NS\omega ^2 \delta a_1 + \frac{N}{8}Sr^2\phi ^2 \delta a_1
- \omega ' \delta {\tilde \omega} + \omega \delta {\tilde \omega }
+ \frac{N}{4} S \phi \delta \xi + \frac{r}{4}
(\phi ' \delta \xi - \phi \delta \xi ')
\label{3-4}
\ee
and
\be
  f'(r^*) = \sigma ^2 \left( -\omega \delta {\tilde \omega } + \frac{r}{4}
\phi \delta \xi \right)
\label{3-5}
\ee
and the Gauss constraint becomes
\be
   \sigma \left( \delta \alpha ' + \omega \delta {\tilde \omega }
- r \frac{\phi}{4} \delta \xi \right) = 0.
\label{3-6}
\ee
Next, we define a `strong' Gauss constraint by
\be
    \delta \alpha ' = -\omega \delta {\tilde \omega }
+ \frac{r}{4}\phi \delta \xi
\label{3-8}
\ee
even when $\sigma = 0$.
\pr
Using (\ref{3-4}) and (\ref{3-8}) we may write
\bea
\delta {\tilde \omega } &=& \frac{r^2}{P}\phi ^2
\left( \frac{\delta \alpha '}{r^2\phi ^2 }\right)'
- \frac{Q}{P} \delta \alpha \nn \\
\delta \xi &=& \frac{4 \omega ^2}{Pr\phi}
\left( \frac{\delta \alpha '}{\omega ^2} \right)'
- \frac{4Q\omega}{Pr\phi} \delta \alpha
\label{3-9}
\eea
where
\bea
    P (r^*) &=& - 2\omega ' + 2 \omega \phi ' + 2 \omega
\frac {N S}{r}  \nn \\
Q(r^*) &=& 2\frac{NS^2}{r^2} \omega ^2 + \frac{1}{4}  NS^2 \phi ^2
- \sigma ^2 .
\label{definPQ}
\eea
If we substitute these expressions into the equation for $\delta {\tilde
\omega}$ we obtain the following equation for $\delta \alpha $:
\bea
   -\delta \alpha ^{{\rm (iv)}} + \left( \frac{2P'}{P} + HP \right)
\delta \alpha ''' & & \nn \\
+ \left\{ \frac{P''}{P} - \frac{2P'^2}{P^2} + 2H'P + Q -\sigma ^2 +
NS^2J
\right\} \delta
\alpha ''  + & & \nn \\
 \left\{ H''P + 2P \left( \frac{Q}{P} \right) ' + \sigma ^2 HP +
\right.  & & \nn \\ \left.
NS^2
\left( -\frac{2\omega P}{r^2}
- \frac{H}{r^2} (\omega^2 - 1) P - \frac{H}{4}\phi ^2 P + \frac{4}{r^2}
\omega ' \right) \right\} \delta \alpha ' + & & \nn \\
 \left\{ P \left( \frac{Q}{P} \right)'' - 2\omega P
 \left( \frac{NS^2}{r^2} \right) ' - \frac{4PNS^2}{r^2}
\omega '+ \sigma ^2 Q - NS^2QJ \right\} \delta \alpha & =& 0
\label{3-10}
\eea
where
\bea
H &\equiv & \frac{2NS}{r P} + \frac{2 \phi '}{\phi P}  \nn \\
J &\equiv & -\frac{2\omega}{r^2} + \frac{\omega ^2 - 1}{r^2}
+ \frac{\phi^2}{4}.
\label{3-11}
\eea
Alternatively, we can eliminate $\delta \xi $ to obtain
the following pair of coupled Schr\"odinger
equations~\cite{brodbeck,eigenmodes}:
\bea
\sigma ^2 \delta \alpha &=& - \delta \alpha '' + \left( \frac{2\phi '}{\phi }
+ \frac{2NS}{r} \right) \delta \alpha ' + \left( \frac{2NS^2}{r^2} \omega ^2
+ \frac{N}{4}S^2\phi ^2 \right) \delta \alpha
+ P\delta {\tilde \omega} \nn \\
\sigma ^2 \delta {\tilde \omega } &=& -\delta {\tilde \omega}''
-\frac{2N}{r^2}S^2 (1 + \omega ) \delta \alpha '
- \left\{ \frac{4\omega ' NS^2}{r^2} + 2\omega (\frac{NS^2}{r^2})' \right\}
\delta \alpha \nn \\
 & &
+ \frac{NS^2}{r^2} \left\{ (\omega - 1)^2
+ \frac{r^2 \phi ^2}{4} \right\}
\delta {\tilde \omega}.
\label{3-12}
\eea
To proceed, it appears necessary to make the following assumptions:
\begin{itemize}
\item
We assume that the equilibrium solutions are continuous functions of the
Higgs mass ${ v}$.

\item
We also assume that given a Schr\"odinger-like equation
$ - \Psi '' + V \Psi = E \Psi $, where the potential $V$
depends continuously on some parameter ${ v}$,
then the bound state energies also depend continuously on ${ v}$.
This can be proven rigorously if
we make the physically plausible assumption of analyticity
of the operators involved in the above system. The proof
then relies on
the powerful
Kato-Rellich theorem of analytic operators~\cite{simon}.
This analyticity requirement is
the case if perturbation theory is used to solve
the Schr\"odinger system with potential $V({ v} + \delta {
v})$ in terms of the spectrum of $V({ v})$, implying that
the changes in the bound state energies $\delta E$ due to
the infinitesimal shift in the parameter ${ v}$
are also infinitesimal. It is also the case where
{\it variational} methods are applicable.
This assumption
implies that the eigenvalues of the discrete spectrum
of the above equations (\ref{3-12}), i.e.
the bound-state energies for $\sigma ^2 < 0 $, are also continuous
as ${ v}$ varies continuously.
\end{itemize}
\pr
The above assumptions had to be made because in the case of
the EYMH black hole system, the solution is not known
analytically but only numerically, and therefore the issue of analyticity
of the various operators involved with respect to
the Higgs field v.e.v., $v$, cannot be rigorously established.
\pr
We now notice that
the continuous spectrum of the equations (\ref{3-12})
is given by $\sigma ^2 > 0$.
Hence, the number of negative modes will change by one
whenever a mode  is either absorbed into the continuum or emerges from it.
\pr
For $\sigma ^2 = 0$ the equations possess pure ``gauge mode" solutions  of the
form
\be
\delta \alpha = \frac{r^2 \Omega '}{2NS^2}, \qquad
\delta {\tilde \omega } = -\omega \Omega, \qquad \delta \xi
= \frac{r \phi }{2} \Omega
\label{3-13}
\ee
where
\be
    \left( \frac{r^2 \Omega '}{2NS^2}\right) '= \left( \omega ^2
+ \frac{r^2}{8}\phi ^2 \right) \Omega.
\label{3-14}
\ee
\pr
Thus, near the event horizon, $\Omega \sim O[(r-r_h)^k] $, where
$k=0$ or $1$, and at infinity,
$\Omega \sim e^{\pm \frac{{ v}}{2}}$
upon choosing $S(\infty ) =1$.
Hence, there is a single non-degenerate, non-normalisable
eigenmode with $\sigma ^2 = 0$.
\pr
For the fourth-order equation with $\sigma ^2 =0$, $\delta \alpha
\sim O[(r-r_h)^k]$ near the horizon, where $k =0$ (twice, corresponding
to the pure ``gauge mode" solutions ) or $k=\frac{-1 \pm \sqrt{5}}{2}$.
\pr
In the latter case,
$\delta {\tilde \omega}, \delta \xi \sim O[(r-r_h)^{k-1}]$, and so
will not remain bounded near the horizon. Hence there is a single
non-degenerate zero mode for non-zero ${ v}$.
\pr
{}From the above it follows that the number of negative modes
of the system (\ref{3-12}) cannot change at any non-zero
value of ${ v}$,
including the bifurcation point ${ v}_{max}$,
by continuity. The negative eigenvalues of this system
are non-degenerate and hence cannot themselves bifurcate at some value
of ${ v}$.
\pr
The only possible place where the number
of negative modes may change is at ${ v} = 0$.
Let $\phi = { v} {\tilde \phi } , \delta \xi = { v} \delta {\tilde
\xi }$. Then the system of equations (\ref{3-12}) becomes
\bea
  -\delta \alpha '' + \left( \frac{2{\tilde \phi } '}{{\tilde \phi }} +
\frac{2NS}{r} \right)  \delta \alpha '
+ \left( \frac{2NS^2{\tilde \phi}^2}{4}\right)  \delta
\alpha & & \nn \\
+ \left( -2\omega ' + 2\omega \frac{{\tilde \phi}'}{{\tilde \phi}}
+ 2\omega \frac{NS}{r} \right) \delta {\tilde \omega}
  & = &  \sigma ^2 \delta \alpha
\nn \\
-\delta {\tilde \omega} '' - 2\frac{NS^2}{r^2}
(1 + \omega) \delta \alpha ' + \frac{NS^2}{r^2}
\left[ (\omega -1)^2 + { v}^2
\frac{r^2{\tilde \phi}^2}{4}\right]
\delta {\tilde \omega} & & \nn \\
- \left[ 4 \omega ' \frac{NS^2}{r^2}
+ 2\omega \left( \frac{NS^2}{r^2}\right) ' \right]  \delta \alpha
 &= & \sigma ^{2} \delta {\tilde \omega}
\label{3-15}
\eea
together with the Gauss constraint
$\delta \alpha ' = -\omega \delta {\tilde \omega} + \frac{1}{4}
{ v}^2 r {\tilde \phi} \delta {\tilde \xi }$.
\pr
Consider the $k=n$ branch of the EYMH solutions. In this branch,
$\frac{{\tilde \phi }'}{{\tilde \phi}}$ has a well-defined limit as
${ v} \rightarrow  0$, and the system (\ref{3-12})
is continuous at ${ v} =0$. In this case,
the Gauss constraint reduces to
$\delta \alpha ' = -\omega \delta {\tilde \omega } $ , and substituting
in (\ref{3-15}) yields the equation
\be
  -\delta \alpha '' + \frac{2}{\omega} \omega ' \delta \alpha '
+ \frac{2NS^2}{r^2}\omega ^2 \delta \alpha =
\sigma ^2 \delta \alpha .
\label{3-16}
\ee
This is the equation studied in ref. \cite{eigenmodes}
where it was shown that there are exactly $n$ negative eigenvalues.
Furthermore, there is a single non-degenerate zero mode given by
\be
\delta \alpha = \frac{r^2 \Omega '} {2NS^2}
\label{3-17}
\ee
where
\be
\left( \frac{r^2 \Omega '}{2NS^2}\right) '=\omega ^2 \Omega .
\ee
\pr
As before, near the horizon
$\Omega \sim O[(r-r_h)] $ or $\Omega \sim O[1]$
whilst at infinity $\Omega \sim r$ or $r^{\frac{1}{2}}$, giving a single
eigenmode with zero eigenvalue.
Thus, the number of negative eigenvalues
does not change at ${ v}=0$
for this branch of solutions.
\pr
For the quasi $k=n-1$ branch of solutions
$\frac{{\tilde \phi}'}{{\tilde \phi}}$ does not
have a well-defined limit as ${ v} \rightarrow 0$, and the
system (\ref{3-12}) is not continuous at ${ v}=0$.
Hence, by continuity, we can conclude that both the $k=n$ and the
quasi $k=n-1$ black holes have {\it exactly} $n$ unstable modes
in the {\it sphaleron} sector.

\section{Instabilities in the Gravitational Sector -
\newline  Catastrophe theory approach}
\pr
It is the aim of this section to prove the
existence and
count the
exact number of unstable modes in the gravitational
sector of the solutions.
In the first part we shall study the conditions
for the existence of unstable modes in a linearized
framework, and we shall study the possibility of a change
in the stability of the system as one varies the Higgs v.e.v.
$v$ continuously from $0$ up to the bifurcation point (c.f. figure 2).
In the second part, which will deal with the change of the stability
of the system at the bifurcation point, we shall go beyond linearized
perturbations by applying catastrophe theory.
It should be stressed that although catastrophe theory was
first employed
by the authors of \cite{torii}, our approach in this section is
somewhat different, and has certain advantages, not least of which
is that we are able to exploit the known stability of the
Schwarzschild black hole to draw conclusions about the non-trivial
EYMH black holes.

\subsection{Linearized perturbations}
\pr
The linearised perturbation equations for the gravitational
sector are:
\bea
-\delta \omega '' + U_{\omega\omega} \delta \omega
+ U_{\omega\phi} \delta \phi &=& \sigma ^2 \delta \omega \nn \\
-\delta \phi '' - \frac{2NS}{r} \delta \phi ' +
U_{\phi\omega} \delta \omega + U_{\phi\phi} \delta \phi &=& \sigma ^2 \delta
\phi
\label{4-1}
\eea
where the prime denotes differentiation with respect to the
tortoise coordinate (\ref{tortoise}) as before, and
\bea
U_{\omega\omega} &=& \frac{NS^2}{r^2} \left[
3\omega ^2 - 1 + \frac{1}{4}r^2\phi^2
- 4r^2 \omega '^2 \left( \frac{N}{r} + \frac{(NS)'}{S}\right) +
\frac{\omega(\omega ^2 -1)}{r} \omega '
+ 2 r \omega ' \phi ^2 \right]
\nn \\
U_{\omega\phi} &=& \frac{NS^2}{r^2}
\left[  \frac{1}{2}(1 + \omega )
\phi r^2 - 2\phi ' \omega ' r^3 \left( \frac{N}{r} + \frac{(NS)'}{S}\right)
+ 2r \phi ' \omega (\omega ^2 - 1) \right.  \nn \\
 & &  \left.
+ \phi \omega '
(1 + \omega )^2 r + \frac{1}{2} \phi ' \phi ^2 (1 + \omega )
+ 2\lambda r^3 \phi \omega ' (\phi ^2 - { v}^2) \right]
\nn \\
U_{\phi\omega} &=& \frac{2}{r^2}U_{\omega\phi} \nn \\
U_{\phi\phi} &=& \frac{NS^2}{r^2} \left[ \frac{1}{2} (1 + \omega )^2 +
\lambda r^2 (3\phi ^2 - {\tilde v}^2)
-2r^3 (\phi ')^2 \left( \frac{N}{r} + \frac{(NS)'}{S} \right)
\right. \nn \\ & &  \left.
+ 2\phi ' \phi (1 + \omega )^2 r + 4 \lambda \phi \phi ' r^3
(\phi ^2 - {\tilde v}^2) \right].
\label{4-2}
\eea
The continuity argument described previously when applied here, implies
that the number of negative eigenvalues of the system (\ref{4-1},\ref{4-2})
can change only when there is a zero mode.
\pr
Suppose that for some ${ v} \ne 0$ there is such a zero mode of
(\ref{4-1}). Given the background solution ($\omega, \phi $)
of the field equations, then $(\omega + \delta \omega, \phi + \delta \phi )$,
together with the corresponding metric functions, will also be a solution
of the field equations.
\pr
As $r \rightarrow \infty $ , then
\be
\left(
\begin{array}{c}
\delta \omega \\
\delta \phi \end{array} \right)
\sim O[e^{-{ v} r^*}]   \qquad r \rightarrow -\infty
\label{arr1}
\ee
and
\be
\left(
\begin{array}{c}
\delta \omega \\
\delta \phi \end{array} \right) \sim {\rm const} \qquad
r \rightarrow -\infty .
\label{ar1}
\ee
Also, since in this sector the change
in the mass function ${m(r)}$ is given by (see Appendix A)
\be
\delta {m(r)} =2N \frac {d\omega }{dr}
 \delta \omega (r) + r^2 N \frac {d\phi }{dr}  \delta \phi
(r)
\label{4-3}
\ee
we have that for this zero mode $\delta {m} \rightarrow 0$
as $r^* \rightarrow \pm \infty $.
Hence, for {\it fixed} parameters $r_h, \lambda, { v}, g$ there
are {\it two solutions} of the original field equations
$\left( \begin{array}{c} \omega \\ \phi \end{array} \right) $,
and
$\left( \begin{array}{c} \omega + \delta \omega \\ \phi
+ \delta \phi \end{array} \right)$,
satisfying the required boundary
conditions. For our purposes we shall {\it assume} that the solution
of ref. \cite{greene} is {\it unique}. Compatibility with the above analysis,
then, requires
the latter to be valid only at the {\it bifurcation point} (c.f. figure 2).
Hence, there is a single zero mode at ${ v}={ v}_{max}$. For any
other ${ v}$ the zero mode is absent.
\pr
Let $\delta \phi \equiv { v}\delta {\tilde \phi }$,
$\phi \equiv { v} {\tilde \phi }$. The equations (\ref{4-2})
become
\bea
-\delta \omega '' + U_{\omega\omega} \delta \omega + { v}^2 {\tilde
U}_{\omega\phi} \delta {\tilde \phi} &=& \sigma ^2 \delta \omega \nn \\
-\delta {\tilde \phi } - \frac{2NS}{r} \delta {\tilde \phi }' +
{\tilde U}_{\phi\omega} \delta \omega + U_{\phi\phi} \delta
{\tilde \phi} &=& \sigma ^2 \delta {\tilde \phi }
\label{4-4}
\eea
where $U_{\omega\phi} \equiv { v} {\tilde U}_{\omega\phi},
{\tilde U}_{\phi\omega} \equiv { v} {\tilde U}_{\phi\omega} $.
\pr
These equations have a well-defined limit as ${ v} \rightarrow 0$,
and are continuous at ${ v} =0$. At ${ v} =0$
the equations reduce to
\be
    \delta \omega '' + U_{\omega\omega} \delta \omega =
\sigma ^2 \delta \omega
\label{4-5}
\ee
where
\be
U_{\omega\omega} = \frac{NS^2}{r^2} \left[ 3\omega ^2 - 1
-4r(\omega ')^2 \left( \frac{N}{r} + \frac{(NS)'}{S} \right) +
\frac{\omega (\omega ^2 - 1)}{r}
\delta \omega ' \right] .
\label{4-6}
\ee
If this equation possesses a zero mode, then $\delta \omega \rightarrow
{\mbox {const}} $, as $r \rightarrow \infty $ for this mode. As $r \rightarrow
\infty $, $N \rightarrow 1 - \frac{{m}}{r} + O[e^{-r}]$,
$\omega \rightarrow -1 + O[e^{-r}]$, $S \rightarrow 1 + O[e^r]$, and the
equation takes the form
\be
\frac{d^2}{d r^2} (\delta \omega ) = -\frac{{m}}{r^2}
\left( 1 - \frac{2 {m}}{r}\right) ^{-1} \frac{d}{dr} (\delta \omega)
+ \frac{2}{r^2} \left( 1 - \frac{{m}}{r}\right) ^{-1} \delta \omega.
\label{4-7}
\ee
Suppose that
$\delta \omega = \sum _{n=0}^{\infty} a_n r^{-\rho - n} $, $a_0 \ne 0$
as $r \rightarrow \infty $. Then $\rho = 2$~or~$-1$.
Let $\delta \omega = r^2 f(r) $. Then, $f(r)$ satisfies the equation
\be
   f'' + f' \left( 1 - \frac{2{m}}{r} \right) ^{-1} \left( \frac{4}{r}
- \frac{6 {m}}{r^2} \right) = 0.
\label{4-8}
\ee
The solution to this equation as $r \rightarrow \infty$ assumes the form
\be
f = \frac{A}{8{\cal M}^3} { \log}\left( \frac{r - 2{\cal M}}{r}
\right)
+ \frac{A}{4{\cal M}^2r^2} + \frac{A}{4{\cal M} r^2} + B
\qquad r \rightarrow \infty
\label{4-9}
\ee
with $A$, $B$ arbitrary constants and ${\cal M}=\lim _{r \rightarrow
\infty  }m(r)$, so that
\be
\delta \omega \rightarrow Br^2 + \frac{A r^2}{8{\cal M}^3}
{\log} \left( \frac{r -
2{\cal M}}{r} \right)
+ \frac{A r}{4{\cal M}^2} + \frac{A}{4 {\cal M}}.
\label{4-10}
\ee
Hence $\delta \omega $ can remain bounded as $r \rightarrow \infty$ only if
$A=B=0$, i.e.  only the trivial solution exists.
Thus, there are no non-trivial zero modes for ${ v} =0$.
\pr
We, therefore, conclude that along each branch of solutions
the number of negative modes remains constant from ${ v}=0$ to
${ v}={ v}_{max}$.

\subsection{Bifurcation points and catastrophe theory}
\pr
In order to determine what happens at ${ v} = { v}_{max}$
we appeal to catastrophe theory~\cite{torii}.
Our aim is to study the possibility of a change of the stability
of the system at $v_{max}$. To this end, we have to determine
a certain function
({\it catastrophe functional}) in the black hole solution which changes
{\it discontinuously} despite the smooth change of certain
(control) parameters of the system.
As we shall show below, in the case at hand the r\^ole
of the catastrophe functional is played by the mass function
of the black hole, whilst the control parameter is the Higgs v.e.v.
$v$. At the bifurcation point $v_{max}$ we shall find
a {\it fold} catastrophe which affects the relevant
stability of the branches of the solution. In addition, the
catastrophe-theoretic approach allows for an {\it exact}
counting of the unstable modes in the various branches.
For notations and mathematical definitions on catastrophe
theory we refer the interested reader to Appendix B.
\pr
We should note at this point that although catastrophe theory
seems powerful enough to yield a universal stability study
of all kinds of non-Abelian black holes~\cite{torii}, however
one should express some caution in drawing conclusions
about absolute stability. Indeed, catastrophe theory
gives information about instabilities of certain modes of the
system. If catastrophe theory gives a stable branch of solution,
this does not mean that the
system is completely stable, given that there
may be other instabilities in sectors where catastrophe theory does not
apply. In our EYMH system this is precisely the case with the sphaleron
sector. However safe conclusions can be reached,
within the framework of catastrophe theory,
regarding {\it relative}
stability of branches of solutions, and it is in this sense that we shall
use it here in order to count the number of unstable modes of the
various branches of the EYMH system.
Having expressed these cautionary remarks we are now ready to proceed
with our catastrophe-theoretic analysis.
\pr
The mass functional ${\cal M}$ (c.f. appendix A)
can be re-written as a functional of the matter fields only as follows:
\pr
\nk First note that ${\cal M}=m(\infty)$. Let
$\mu (r) \equiv m(r) - m(r_h) = m(r) - \frac{r_h}{2}$.
Then, using a prime to denote $d/dr $
\bea
 \mu '(r) = m'(r)  &=& \frac{1}{2}
\left[ \left( 1-\frac{2{m}}{r}\right) (2(\omega ')^2
+ r^2 (\phi ')^2 ) \right] \nn \\ & &
+ \frac {r^2}{2} \left[ \frac{(1-\omega ^2)^2}{r^4}
+ \frac{\phi ^2}{2r^2} (1 + \omega )^2 +
\frac{\lambda }{2} (\phi ^2 - { v}^2 )^2  \right]  \nn \\
 & = & \frac{1}{2}
\left[ \left(1 - \frac{r_h}{r} \right) (2 (\omega ')^2 + r^2 (\phi ')^2 )
\right] \nn \\ & &
+ \frac {r^2 }{2}  \left[ \frac{(1-\omega ^2)^2}{r^4} +
\frac{\phi ^2 (1 + \omega )^2 }{2r^2} +
\frac{\lambda}{2}(\phi ^2 - {\tilde v}^2)^2 \right] \nn \\
& &
- \frac{\mu}{r} (2 (\omega ')^2 + r^2 (\phi ')^2 ).
\label{4-11}
\eea
The last term on the right-hand-side can be written in terms of the
metric function $\delta $ (cf appendix A)
\be
      -\frac{\mu}{r} (2(\omega ')^2 + r^2 (\phi ')^2 ) = \mu \delta '.
\label{4-12}
\ee
Solving for $\mu$ gives
\be
 \mu (r) =e^{\delta (r)} \int _{r_h} ^r {\cal K} [\omega, \phi ]
e^{-\delta (r')} dr'
\label{4-13}
\ee
where
\bea
{\cal K}[\omega,\phi] &  \equiv & \frac{1}{2} \left[ \left(1
- \frac{r_h}{r}\right)
(2(\omega ')^2 + r^2 (\phi ')^2 ) \right] \nn \\
 & &
+ \frac {r^2}{2} \left\{ \frac{(1- \omega ^2)^2}{r^4} + \frac{\phi ^2
(1 + \omega )^2 }{2r^2}
+ \frac{\lambda}{2} (\phi ^2 - { v}^2 )^2 \right\}.
\label{4-14}
\eea
Hence, setting $\delta (\infty ) = 0$ we obtain
\be
{\cal M} = \frac{r_h}{2} +
\int _{r_h}^\infty {\cal K}[\omega, \phi] e^{-\delta
(r)} dr.
\label{4-15}
\ee
Varying this functional with respect to the matter fields
yields the correct equations of motion~\cite{heuslerstr}.
Thus, the equilibrium solutions of the field equations
will be stationary points of the functional ${\cal M}$.
\pr
If we plot the solution curve in $({ v}, \delta _0, {\cal M} )$
space, then the resulting curve is smooth  (c.f. figure 1).
For the {\it Catastrophe Theory} (c.f. Appendix B) we consider
$\delta _0$ as a variable, and ${ v}$
as a {\it control } parameter. The Whitney surface
could be defined in our case as follows.
For each ${ v}$, consider a smoothly varying set of functions
$\omega _\delta, \phi _\delta $ indexed by the value of $\delta _0$ they give:
\be
    \delta _0 = \int _{r_h}^\infty \frac{1}{r} (2 (\omega _\delta ')^2
+ r^2 (\phi _\delta ')^2 ) dr
\label{4-16}
\ee
such that $\omega _\delta ,  \phi _\delta  $ are the appropriate solutions
to the field equations when $\delta _0$ lies on the solution curve.
Then, the solution curve represents the curve of extremal points of
this Whitney surface, $\delta {\cal M} =0$.
\pr
The projections of this curve onto the $({ v}, \delta _0 )$
and $(\delta _0, {\cal M} )$ planes are also smooth curves.
The catastrophe map $\chi $ projects the solution curve onto the
$({ v}, {\cal M})$ plane
\be
  \chi~:~({ v},\delta _0, {\cal M} ) \rightarrow
({ v}, {\cal M}).
\label{4-17}
\ee
This yields the curve shown in figure 2. This map is regular except at the
point $({ v}={ v}_{max}, {\cal M}={\cal M}_{max} )$,
where it is singular. This point is the {\it bifurcation set} B.
\pr
Since the Whitney surface describes a one-parameter (${ v})$
family of functions of a single variable ($\delta $), and the bifurcation
point is a single point, we have a fold catastrophe, as found in ref.
\cite{torii} from a different point of view. A more detailed comparison,
of the results of that reference with ours will be made at the end of the
section.
\pr
Catastrophe theory tells us that the stability of the system
will change at the point B on the solution curve, and, furthermore,
that the branch of solutions (including the point B) having the higher
mass (for the same value of ${ v}$) will be
{\it unstable}, relative to the other branch. Hence,
from our previous continuity considerations, the $k=n$ branch
of solutions will have {\it exactly one more} negative mode than the
quasi- $k=n-1$ branch.
\pr
The catastrophe theory analysis applies to the gravitational sector
rather than the  sphaleronic sector, since gravitational perturbations
correspond to {\it small} changes in the functions
$\omega $ and $\phi $, whilst keeping the functional form of
${\cal M}$ {\it fixed}. On the contrary, sphaleronic perturbations
keep the functions $\omega$ and $\phi $ fixed, affecting the functional
form of ${\cal M}$. As we discussed in the previous section, the number
of unstable sphaleron modes  is the {\it same} for the $k=n$ and
quasi-$k=n-1$ branches.
\pr
All that remains is to determine the number of
negative modes of the quasi-$k=0$ branch of solutions.
{}From the above considerations, this will be equal to the number
of negative modes of the ${ v}=0$ limiting case of this branch
of solutions, which is nothing other than the Schwarzschild black hole.
The gravitational perturbation equation is in this case,
where a prime now denotes $d/dr^{*}$,
\be
-\delta \omega '' + U_{\omega\omega} \delta \omega = \sigma ^2 \delta \omega
\label{4-18}
\ee
where
\be
U_{\omega\omega} =\frac{NS^2}{r^2}
\left[ 3\omega ^2 - 1 - 4r(\omega ')^2
\left( \frac{N}{r} + \frac{(NS)'}{S} \right)
+ \frac{\omega (\omega ^2 - 1)}{r}
\delta
\omega ' \right].
\label{4-19}
\ee
For the Schwarzschild solution $N=1 -\frac{r_h}{r}, S=1, \omega = 1$.
Equation (\ref{4-18}), then, reduces to
\be
-\delta \omega '' + \frac{2}{r^2} \left( 1 -\frac{r_h}{r}
\right) \delta \omega =
\sigma ^2 \delta \omega
\label{4-20}
\ee
which has the form of a standard one-dimensional Schr\"odinger
equation with potential
\be
   V(r^*) =\frac{2}{r^2} \left( 1 -\frac{r_h}{r} \right) \qquad
\frac{d}{dr ^* } \equiv \left( 1 - \frac{r_h}{r} \right) \frac {d}{dr}.
\label{4-21}
\ee
As $r^* \rightarrow \pm \infty $, $V(r^*) \rightarrow 0$.
On the other hand, for finite $r^*$, $-\infty < r ^* < \infty $,
the potential is positive definite, $V(r^*) > 0$.
\pr
Then, by a standard theorem of quantum mechanics~\cite{messiah},
the Schr\"odinger equation (\ref{4-20},\ref{4-21})
has no bound states. Thus, the Schwarzschild solution (and, hence, the
quasi-$k=0$ branch of solutions) has no negative gravitational
modes, a known result in agreement with the no-hair conjecture.
\pr
Working inductively through the various branches of solutions
(the ${ v} = 0$ limit of the
$k=n$ branch is the same as that for the quasi-$k=n-1$
branch after replacing $\omega $ by $-\omega$) we find that
the $k=n$ branch possesses exactly $n$ unstable gravitational
modes, and the quasi-$k=n-1$ branch exactly $n-1$
negative modes. This result has been conjectured but not proven
in ref. \cite{lavmaison}.
\pr
Before closing the section we would like to compare our
results with those of ref. \cite{torii}.
In ref. \cite{torii} the authors also used catastrophe theory to draw
conclusions about the stability of the EYMH black holes, but their
approach was somewhat different.
There the writers fixed the parameter $\lambda =0.125$ and also
fixed the Higgs mass $v$.
They then varied the horizon radius $r_{h}$ and for each solution
calculated the value of the mass functional $\cal M$, the field
strength at the horizon $B_{h}$ given by
\be
B_{h}=|F^{2}|^{\frac {1}{2}} \left| _{\mbox {horizon}} \right.
\ee
and the Hawking-Bekenstein entropy $S=\pi r_{h}$, to give
a smooth solution curve in $({\cal M},B_{h},S)$ space.
The projection of this curve on to the $({\cal M},S)$ plane
has the same qualitative features as our figure 2.
Here we have also fixed $\lambda =0.15$, and in addition we have
fixed $r_{h}=1$ and varied the Higgs mass $v$ from 0 up to the
bifurcation point.
Torii et al concluded that the $k=1$ branch of solutions was more
unstable than the quasi-$k=0$ branch of solutions.
The advantage of our approach is that, by interpolating between
the various coloured black hole solutions \cite{bizon}, beginning
with the Schwarzschild solution, we will be able to calculate the
exact number of unstable modes of each branch of solutions and not
just give qualitative information concerning their relative
stability.

\section{Entropy considerations}

It remains now to associate the above catastrophic considerations
with some elementary `thermodynamic' properties of the black hole
solutions, and in particular their entropy.
Our aim in this section is to give elementary
estimates of the entropy of the various branches, assuming
thermodynamic equilibrium of the black hole with a surrounding
heat bath. Such estimates will allow an association of the
stability issues with the amount of entropy carried by the various branches
of the solution. In particular we shall argue that the `high-entropy'
branch has relatively fewer unstable modes than the `low-energy' ones
and, thus, is relatively more stable.
We shall employ approximate WKB semi-classical methods
for the evaluation of the entropy. We shall also study the conditions
under which such estimates are valid.
\pr
The calculation of the entropy
of the black hole will be made on the basis
of calculating the entropy
of quantum matter fields
in the black hole space time.
This will constitute only a partial
contribution to the total black hole entropy.
A complete calculation requires a proper quantization
of the gravitational field, which at present is not possible,
given the non-renormalizability of (local) quantum gravity.
Ignoring such back reaction effects of the matter fields
to the (quantum) geometry of space time results in ultra-violet
divergences in the calculated entropy of the matter
fields~\cite{thooft,susskind}.
Such divergences can be absorbed
in a renormalization of the gravitational (Newton's) constant.
This is so because the entropy is proportional
to the area of the black hole horizon,
with the divergent contributions appearing as
multiplicative factors.
\pr
In what follows we shall estimate the entropy of a scalar field
propagating in the EYMH black hole background.
Anticipating a path integral formalism
for quantum gravity, we shall compute only the entropy
which is due to quantum fluctuations of the scalar field
in the black hole background.
The part that contains the classical solutions to the equations
of motion contributes to the `classical' entropy, associated
with the classical geometry of space time. This part is known
to be proportional to  $1/4$ of the horizon area~\cite{thooft,susskind}.
The quantum-scalar-field entropy part will also turn out to be
proportional to the horizon area, but the
proportionality coefficient is linearly
divergent as the ultraviolet cut-off is removed, exactly as it happens
in the corresponding computation for the Schwarzschild
black hole~\cite{thooft}.
Absorbing the divergence into a conjectured renormalization
of the gravitational constant will enable us to estimate the entropy
of the various branches of the EYMH black hole solution, and relate this
to the above-mentioned catastrophe-theoretic arguments.
\pr
As we shall show, this will be possible only in the {\it non-extremal} case,
which is the case of the numerical solutions studied in ref. \cite{greene}
and in the present work. Among the solutions, however,
there exist some {\it extremal} cases, for which the Hawking temperature -
which is defined by assuming thermal equilibrium
of the black hole system with a surrounding heat bath - vanishes.
In such a case, the linearly divergent entropy of the scalar field
vanishes. However, there are non trivial {\it logarithmically}
divergent contributions to the black hole entropy
which cannot be absorbed in a renormalization of the gravitational
constant. Moreover, the classical Bekenstein-Hawking entropy formula
seems to be violated by such contributions to the
black-hole entropy.
The situation is similar to the case of a scalar field
in an extreme (3+1)-dimensional
dilatonic black hole background~\cite{dilatonbh}, and seems to be
generic to black holes with non-conventional hair.
We shall briefly comment on this issue at the end of the section.
\pr
We shall be brief in our discussion and concentrate only in basic new results,
relevant for our discussion above. For details in the formalism we refer
the interested reader to the existing literature~\cite{thooft,susskind}.
To start with, we note that
the metric for the EYMH black holes is given by:
\be
ds^{2} =
-\left( 1- \frac{2m(r)}{r} \right) e^{-2\delta (r)} dt^{2}
+ \left( 1- \frac{2m(r)}{r} \right) ^{-1} dr^{2}
+r^{2}( d \theta ^{2} + \sin ^{2} \theta d\varphi ^{2}).
\ee
Consider a scalar field of mass $\mu $ propagating in this spacetime
\cite{thooft},
satisfying the Klein-Gordon equation:
\be
\frac {1}{\sqrt{-g}} \partial _{\mu } ( \sqrt {-g}
g^{\mu \nu } \partial _{\nu} \Phi )
- \mu ^{2} \Phi =0.
\ee
Since the metric is spherically symmetric, consider solutions of the
wave equation of the form
\be
\Phi (t,r,\theta, \varphi)= e^{-iEt} f_{El}(r) Y_{lm_{l}}(\theta,\varphi)
\ee
where $Y_{lm_{l}}(\theta ,\varphi )$ is a spherical harmonic
and $E$ is the energy of the wave.
The wave equation separates to give the following radial equation for
$f_{El}(r)$
\bea
\left( 1-\frac{2m(r)}{r} \right) ^{-1} E^{2} f_{El} (r) & & \nn \\
+ \frac {e^{\delta (r)}}{r^{2}} \frac {d}{dr}
\left[ e^{-\delta (r)} r^{2} \left( 1- \frac {2m(r)}{r} \right)
\frac {df_{El} (r)}{dr} \right] & & \nn \\
-\left[ \frac {l(l+1)}{r^{2}} +\mu ^{2} \right] f_{El} (r) & = & 0.
\eea
The ``brick wall'' boundary condition is assumed \cite{thooft},
 namely,  the wave
function is cut off just outside the horizon,
\be
\Phi =0 \mbox { at } r=r_{h} + \epsilon
\ee
where $r_{h}$ is the black hole horizon radius, and $\epsilon$ is
a small, positive, fixed distance which will play the r\^ole
of an ultraviolet cut-off. We
also impose an infra-red cut-off at a very large distance $L$ from
the horizon:
\be
\Phi =0 \mbox { at } r=L, \mbox { where } L \gg r_{h}.
\ee
Hence $f$ satisfies
\be
f_{El} (r) =0 \mbox { when } r=r_{h} + \epsilon
\mbox { or } r=L.
\ee
In anticipation of being able to use a WKB approximation, define
functions $ K(r)$ and $h(r)$ by
\bea
K^{2} (r,l,E) &  = & \left( 1- \frac {2m(r)}{r} \right) ^{-1}
\left[ E^{2} \left( 1- \frac {2m(r)}{r} \right) ^{-1}
- \frac {l(l+1)}{r^{2}} - \mu^{2} \right] \label{waveno} \\
h(r) & = & e^{-\delta (r)} r^{2} \left(
1-\frac {2m(r)}{r} \right) .
\eea
Then the equation for $f_{El}(r)$ becomes
\be
\frac {1}{h(r)} \frac {d}{dr} \left[
h(r) \frac {d}{dr} f_{El} (r) \right] +
K^{2} (r,l,E) f_{El} (r) =0.
\ee
Now define a function $u(r)$ by
\be
f_{El} (r)= \frac {u(r)}{\sqrt {h(r)}}.
\ee
Then $u(r)$ satisfies
\be
\frac {d^{2}u}{dr^{2}} +\left[
K^{2} +\frac {1}{4h^{2}} \left( \frac {dh}{dr} \right) ^{2}
-\frac {1}{2h} \frac {d^{2}h}{dr^{2}} \right] u =0.
\ee
The WKB approximation for $u$ will be valid if
\be
\left| \frac {1}{4h^{2}} \left( \frac {dh}{dr} \right) ^{2}
-\frac {1}{2h} \frac {d^{2}h}{dr^{2}} \right|
\ll
\left| K^{2} \right|
\ee
and
\be
\left| \frac {dK}{dr} \right| \ll \left| K^{2} \right| .
\ee
The first inequality is required so that $u$ can be taken to satisfy
the equation
\be
\frac {d^{2}u}{dr^{2}} +K^{2}u =0
\ee
where $K$ is now the radial wave number, and the second inequality is
required so that the approximation to the wave function
\be
u(r) \sim \frac {1}{\sqrt {K(r)}} \exp
\left[ \pm i \int K(r) dr \right]
\ee
is valid.  Assuming, for the present, that the WKB approximation
is valid, define the radial wave-number $K$ as above whenever the
right-hand-side of (\ref{waveno}) is non-negative.
Define $K^{2}=0$ otherwise.
Then the number of radial modes $n_{K}$ is given by
\be
\pi n_{K} = \int _{r_{h}+\epsilon } ^{L}
dr K(r,l,E) \label{nk}
\ee
where the fact that $n_{K}$ must be an integer restricts the possible
values of $E$.
For fixed energy $E$, the total number $N$ of solutions with
energy less than or equal to $E$ is
\bea
\pi N & = & \int (2l+1) \pi n_{K} dl \nn  \\
 & = &
\int _{r_{h}+\epsilon }^{L}
\left( 1-\frac {2m(r)}{r} \right) ^{-1} dr
\int (2l+1) dl  \nn \\ & & \times
\left[ E^{2} -\left( \frac {l(l+1)}{r^{2}} + \mu ^{2}
\right) \left( 1-\frac {2m(r)}{r} \right) \right]
^{\frac {1}{2}}
\eea
where the integration is performed over all values of $l$ such that
the argument of the square root is positive.
\pr
The Hawking temperature of the black hole is given by
\be
T^{-1}=\beta =
\frac {4\pi r_{h} e^{\delta _{h}}}{1-2m'_{h}} \label{temp}
\ee
where $\delta _{h}$ is fixed by the requirement that
$\delta (\infty ) =0$, and $'=d/dr $.
Assume further that $\beta^{-1} \ll 1 $.
It should be noted that $T=0$ in the extreme case $m'_{h}=0.5$.
Further comments on the entropy in this situation will be
made at the end of the section.
However, as discussed in  Appendix A, this situation does not arise
for the black holes we are concerned with.
The free energy $F$ of the system is given by
\bea
e^{-\beta F}  &=&
\sum e^{-\beta E} \nn \\
 & = &
\prod _{n_{K},l,m_{l}} \frac {1}{1-\exp (-\beta E)}.
\eea
Hence
\bea
 \beta F & = &
\sum _{n_{K},l,m_{l}} \log (1- e^{-\beta E}) \nn  \\
 & \simeq &
\int dl (2l+1) \int dn_{K} \log (1-e^{-\beta E})
\eea
for large $\beta $, integrating over appropriate $l$, $E$.
Integrating by parts,
\bea
F & = &
-\frac {1}{\beta } \int dl (2l+1) \int d(\beta E)
\frac {n_{K}}{\exp (\beta E) -1} \nn \\
 & = & -\frac {1}{\pi} \int dl (2l+1) \int dE
\frac {1}{\exp (\beta E) -1} \int _{r_{h}+\epsilon }^{L}
dr \nn \\ & & \times
\left(1-\frac {2m(r)}{r} \right) ^{-1} \left[
E^{2} -\left( 1-\frac {2m(r)}{r} \right) \left(
\frac {l(l+1)}{r^{2}} +\mu ^{2} \right) \right] ^{\frac {1}{2}}
\eea
where we have substituted for $n_{K}$ from (\ref{nk}).
The $l$ integration can be performed explicitly,
\bea & &
\int dl (2l+1) \left[ E^{2} \left( 1-\frac {2m(r)}{r} \right)
\left( \frac {l(l+1)}{r^{2}} + \mu ^{2} \right)
\right] ^{\frac{1}{2}} \nn \\
 &= & \frac  {2}{3} r^{2} \left( 1-\frac {2m(r)}{r} \right) ^{-1}
\left( E^{2} -\left( 1-\frac {2m(r)}{r} \right) \mu^{2}
\right) ^{\frac {3}{2}}
\eea
to give
\be
F= -\frac {2}{3\pi} \int dE \frac {1}{\exp (\beta E)-1}
\int _{r_{h}+\epsilon }^{L} dr r^{2}
\left( 1-\frac {2m(r)}{r} \right) ^{-2}
\left[ E^{2} -\left( 1-\frac {2m(r)}{r} \right) \mu^{2}
\right] ^{\frac {3}{2}}.
\ee
Introduce a dimensionless radial co-ordinate $x$ by
\be
x=\frac {r}{r_{h}}.
\ee
Then
\be
F= -\frac {2r_{h}^{3}}{3\pi} \int dE
\frac {1}{\exp (\beta E) -1} \int _{1+{\hat {\epsilon }}}^{\hat {L}}
dx x^{2} \left( 1-\frac {2{\hat{m}}(x)}{x} \right) ^{-2}
\left[ E^{2} - \left (1-\frac {2{\hat {m}}(x)}{x} \right)
\mu ^{2} \right] ^{\frac {3}{2}} \label{integrand}
\ee
where ${\hat {\epsilon }} = \frac {\epsilon }{r_{h}}$,
${\hat {L}} =\frac {L}{r_{h}} $,
and
${\hat {m}}(x)=\frac {m(xr_{h})}{r_{h}} $.
\pr
The contribution to $F$ for large values of $x$ is
\be
F_{0}=-\frac {2}{9\pi} L^{3}
\int _{\mu}^{\infty} dE \frac {(E^{2}-\mu^{2})^{\frac {3}{2}}}
{\exp (\beta E) -1}
\ee
which is the expression for the free energy in flat space.
The contribution for $x$ near $1$ diverges as $\epsilon \rightarrow
0$.
For $x$ near $1$, the leading order term in the integrand
in (\ref{integrand}) is
\be
E^{3}(x-1)^{-2}(1-2{\hat {m}}'_{h})^{2}
\ee
where
${\hat {m}}'_{h} = {\hat {m}}' (1)= m'(r_{h})$.
This gives the leading order divergence in $F$
\bea
F_{div} & = &
-\frac {2r_{h}^{3}}{3\pi}
\frac {(1-2{\hat {m}}'_{h})^{-2}}{\hat {\epsilon }}
\int dE \frac {E^{3}}{\exp (\beta E) -1} \nn \\
 & = &
-\frac {2\pi ^{3}}{45 {\hat {\epsilon }}}
\frac {r_{h}^{3} (1-{\hat {m}}'_{h} )^{-2}}{\beta ^{4}}
\nn  \\
 & = &
-\frac {2\pi ^{3}}{45 \epsilon }
\frac {r_{h}^{4} (1-{\hat {m}}'_{h}) ^{-2}}{\beta ^{4}}.
\eea
The total energy $U$ and entropy $S$ are given by
\be
U= \frac {\partial }{\partial \beta }(\beta F)
= \frac {2\pi ^{3}}{15 \epsilon }
\frac {r_{h}^{4} (1-2 {\hat {m}}'_{h})^{-2}}{\beta ^{4}}
\ee
\be
S=\beta ^{2} \frac {\partial F}{\partial \beta }
= \frac {8\pi ^{3}}{45 \epsilon }
\frac {r_{h}^{4} (1-2{\hat {m}}'_{h} )^{-2}}{\beta ^{3}}.
\label{entrop2}
\ee
Substituting for $\beta $ from (\ref{temp}), obtain
\be
S=\frac {r_{h}}{360 \epsilon} (1-2m_{h}' ) e^{-3\delta _{h}}.
\label{entropy}
\ee
\pr
Before discussing the implications of this formula, it is
necessary to ascertain when the approximations used are valid.
Firstly, $ \beta \gg 1$ if $r_{h} \gg 1 $ or $1-2m_{h}' \ll 1$.
In the first case, non-trivial (viz. non-Schwarzschild) solutions
exist only for very small values of $v $, the Higgs mass \cite{torii}
and these solutions will be very close to the Schwarzschild
solution having mass $M=r_{h}/2$.
In the second case, the black hole is very nearly extremal. For
$1=2m_{h}'$ exactly, the above analysis does not apply.
However, $1-2m_{h}' \ll 1$ for large $n$ \cite{bizon} and for
any value of $v $ for which a non-trivial solution exists.
We shall discuss the physical implications of this (nearly)
extremal case at the end of the section.
\pr
Secondly, consider the validity of the WKB approximation.
The principal contribution to the free energy $F$ comes from
the region where $K$ is large; in particular, above we have
concentrated on $x$ close to 1.
It is expected that the WKB approximation will be valid
when $K$ is large.
For $K$ large, it may be approximated by
\be
K=E\left( 1-\frac {2m(r)}{r} \right) ^{-1}.
\ee
For $r$ near $r_{h}$, then
\be
K=E(1-2m_{h}')^{-1}(r-r_{h})^{-1} + 0(1)
\ee
whence
\be
\frac {dK}{dr} =-E(1-2m_{h}')^{-1}(r-r_{h})^{-2}.
\ee
Hence
\be
\left| \frac {dK}{dr} \right| \ll |K^{2}|
{\mbox { if }}
\left| \frac {E}{1-2m'_{h}} \right| \gg 1.
\ee
Similarly, for $r$ near $r_{h}$ we may approximate $h$ by
\be
h=e^{-\delta }r^{2}(1-2m_{h}')(r-r_{h}) + 0(r-r_{h})^{2}.
\ee
Then
\be
\frac {1}{4h^{2}} \left( \frac {dh}{dr} \right) ^{2}
-\frac {1}{2h} \left( \frac {d^{2}h}{dr^{2}} \right)
= \frac {1}{4(r-r_{h})^{2}} + 0(r-r_{h})^{-1},
\ee
so that
\be
\left| \frac {1}{4h^{2}} \left( \frac {dh}{dr} \right) ^{2}
-\frac {1}{2h} \frac {d^{2}h}{dr^{2}} \right|
\ll |K^{2}|
\mbox { if } \left| \frac {4E}{1-2m_{h}'} \right| \gg 1.
\ee
Thus, for black hole solutions with large $n$, the WKB  approximation
is valid except for small values of $E$.
Now return to the expression for the entropy (\ref{entrop2}),
\be
S \equiv S_{linear}
= \frac {r_{h}}{360 \epsilon } (1-2m_{h}') e^{-3\delta _{h}}.
\label{linearen}
\ee
We notice first that the entropy is positive, due to the fact that
for the solutions $m_h' < 1/2$ to avoid naked singularities.
Having said that, we now
fix $n$ and consider the two branches of black hole solutions, the
$k=n$ and quasi-$k=n-1$ solutions.
The linear divergence $r_h/\epsilon $ is a common multiplicative factor
in all branches, and thus can be absorbed in a renormalization of the
gravitational constant~\cite{susskind}.
This can be done as follows: Re-write $r_h/\epsilon = 4\pi r_h^2
\frac{1}{\epsilon 4\pi r_h}$,
where $A=4\pi r_h^2$ is the horizon area,
$G_0$ is the bare graviational coupling constant
(which,
by convention, had been set to one in the
previous formulae),
and
$r_{h}\epsilon=2m_{h} \epsilon$
may be considered as the invariant distance (cut-off)
of the brick wall from the horizon.
The classical Bekenstein-Hawking entropy formula
is then still valid, but with the renormalized gravitational constant $G_R$
replacing the bare (classical) one $G_0$
\be
       S_{classical} + S_{linear} = (\frac{1}{4G_0} + O[\frac{1}{\epsilon}])A
=\frac{1}{4G_R}A
\label{beh}
\ee
\pr
Such a renormalization
may be thought of as expressing quantum matter back reaction effects
to the space-time geometry.
Doing this in our case, we observe from (\ref{linearen}) that
for each $v$, the $k=n$ solution has larger $m_{h}'$ and
$\delta _{h}$ than the quasi-$k=n-1$ solution.
Hence the $k=n$ solution has a lower entropy than the
quasi-$k=n-1$ solution, in agreement with Torii et al \cite{torii}.
\pr
Before closing the section we would like to make some
important comments
concerning the extreme case $m'(r_h)=1/2$, for which the Hawking
temperature (\ref{temp}) {\it vanishes}.
In this case the linearly-divergent part of the entropy
(\ref{entropy}) also vanishes, but this is not the case
for the next-to-leading order {\it logarithmically} divergent
part\footnote{It should be noted that the
logarithmic divergent parts exist also in the non-extreme case, but
there they are suppressed by the dominant linearly divergent terms.
It can be easily checked that for the solutions of ref. \cite{greene},
their presence does not affect the entropy considerations above, based
on the linearly divergent term.}.
\pr
The logarithmic divergent part of the free energy can be found
from (\ref{integrand}) by requiring the following
expansion
\be
   {\hat m}(x) = {\hat m}_h + {\hat m}'_h (x-1) +
\frac{1}{2} {\hat m}_h'' (x-1)^2 + \dots  \qquad {\hat m}_h = \frac{1}{2}.
\label{expn}
\ee
Using the trick $x = 1 + (x-1)$ we can write down the identities:
\bea
  x^{-1} &=& 1 - (x-1) + (x-1)^2 + \dots \nn \\
x^{-2} &=& 1 + 2 (x-1) + (x-1)^2.
\label{ident}
\eea
Hence
\bea
  1-\frac{2{\hat m}(x)}{x} &=& (1 - 2{\hat m}'_h)(x-1) +
(2{\hat m}_h - 1 - {\hat m}_h'') (x-2)^2 + \dots     \nn \\
\left( 1-\frac{2{\hat m}(x)}{x} \right) ^{-2} &=& (1-2{\hat m}'_h)^{-2}
(x-2)^{-2} \nn \\ & & \times
\left\{ 1 + \frac{2(x-1)
(2{\hat m}'_h - 1- {\hat m}''_h )}{2{\hat m}'_h -1}  +\dots \right\}.
\label{comput}
\eea
Substituting in (\ref{integrand}) we obtain for the next-to-leading
order divergence of the free energy
\bea
F_{nlo} &=&   \frac{2r^3_h}{3\pi} \frac{2 (2-4{\hat m}'_h + {\hat m}''_h )}
{(1 - {\hat m}'_h)^3} \int dE \frac{E^3}{e^{\beta E} - 1} \int _{1 + {\hat
\epsilon}} dx \frac{1}{x-1}  \nn \\ & & +
\frac{2r_h^3}{3\pi} \frac{3}{2}\mu^2 \frac{1}{1-2{\hat m}'_h}
\int dE \frac{E}{e^{\beta E}- 1} \int _{1 + {\hat \epsilon}} \frac{dx}{x-1}.
\label{intgr}
\eea
Using the formulae
\bea
  \int _0^\infty dE \frac{E^3}{e^{\beta E} - 1} &=& \frac{\pi ^4}
{15\beta ^4}
\nn \\
\int _0^\infty dE \frac{E}{e^{\beta E} -1 } &=& \frac{\pi ^2}{6 \beta ^2}
\label{resintgr}
\eea
the expression (\ref{intgr})
reduces to
\be
F_{nlo} = \frac{4}{45}r_h^3 \frac{\pi ^4}{\beta ^4}
\frac{2 - 4 {\hat m}'_h + {\hat m}''_h }{(1 - 2{\hat m}'_h )^3 }
\log {\hat \epsilon} - \frac{1}{6} r_h^3 \frac{\pi}{\beta ^2}
\mu ^2 \frac{1}{1- 2{\hat m}'_h }\log {\hat \epsilon}.
\label{nloF}
\ee
{}From (\ref{entrop2}) the corresponding
next-to-leading contribution to the entropy in the extremal case
${\hat m}'_h =1/2$
(where the linear divergence vanishes) is given by
the following expression :
\be
  S_{nlo} = \left[ \frac{1}{3}r_h^2 \mu ^2 e^{-\delta _h}
- \frac{1}{180} e^{- 3\delta _h} {\hat m}''_h \right]
\log \left( \frac{\epsilon}{r_h} \right)
\label{entrop3}
\ee
Thus, we observe that in the extremal case the entropy
diverges logarithmically  with the ultraviolet cut-off,
in a similar spirit to the case of the dilatonic black hole
background~\cite{dilatonbh}. In our case, however, the
horizon area does not vanish, because there is no dilaton
field exponentially coupled to the graviton.
Thus, one could hope that the divergent contribution
(\ref{entrop3}) could be absorbed in the renormalization
of the gravitational constant, so that a formal
Bekenstein-Hawking expression for the
entropy is still valid. However,
as we see from (\ref{entrop3}), for generic
scalar fields this cannot be the case,
due to terms
that spoil the proportionality
of $S_{nlo}$ to the black hole
horizon area $A=4\pi r_h^2$.
Indeed, let us analyze the various contributions
in (\ref{entrop3}).
\pr
The first term can be absorbed into a renormalization
of the gravitational constant, and respects the
classical formula (\ref{beh}). This is not the case with the
second term however. From
equation (\ref{double}) of Appendix A, we observe that
there are contributions that depend on the (boundary) horizon values
of the fields $\phi _h$ and $\omega _h$
which are not proportional to  the horizon area $A$,
\be
     m_h'' =\frac{1}{r_h} \left( 1- \frac{\phi _h ^2}{2\omega _{h}}
(1+\omega _{h})^2 \right)
\label{curious}
\ee
Thus, the associated contribution to the black hole entropy
seems not to be related to geometric aspects of the black hole
background.
\pr
One is tempted to interpret
such contributions
as being associated with information loss across the horizon.
This is supported by the fact that the logarithmic divergencies
disappear for black holes whose horizon is vanishing in the sense of
$r_h \rightarrow \epsilon$.
For consistency with
the interpretation as loss of information,
the {\it positivity} requirement of the relevant contribution
to the black hole entropy has to be imposed.
Returning to formula (\ref{entrop3}) we observe that
the above requirement implies $m_h'' > 0$.
In the present case we do not know whether
extremal solutions exist. {\it A priori} there is no
reason why such solutions should not exist in the EYMH system.
If such a solution exists, the above-mentioned positivity
requirement will impose restrictions on the
boundary (horizon) values of the hair fields of the black hole
background.
{}From the case at hand, it seems that the ambiguities
in sign are associated with the presence of the non-abelian
gauge field component $\omega$. Indeed, from (\ref{curious}),
it is immediately seen that the contributions of the scalar Higgs field
$\phi $ alone
to $m_h''$ are manifestly positive. The terms that
could lead to negative logarithmic contributions to the entropy
are associated with the field $\omega$ and vanish for $\omega = -1$.
\pr
This phenomenon is somewhat similar to what is happening
in the case
of a (spin one) gauge field in the presence of an
ordinary black hole background.
If one integrates quantum fluctuations
of a spin one field in a gravity background, there is an induced
coefficient in front of the Einstein curvature term in the
effective action whose sign is negative for space-time dimensions less than
$8$~\cite{kabat}.
Notice that such sign ambiguities do not occur for scalar fields
in conventional black hole backgrounds.
In our case,
there are gauge fields present in the black hole
background associated with non-conventional hair.
The sign ambiguities found above in the logarithmically-divergent
contributions to the entropy (\ref{entrop3})
occur already when one considers
quantum fluctuations of
scalar fields. This is associated with negative signatures of terms
that involve the gauge field hair background in the effective action.
\pr
Some comments are now in order concerning the
the so-called {\it entanglement} entropy~\cite{bombelli}
of fields in background space times with event horizons or
other space-time boundaries.
The entanglement entropy is obtained from the density matrix of the
field upon tracing over degrees of freedom that cross the event horizon
or lie in the interior of the black hole, and therefore is
closely associated with loss of information. The entanglement entropy
is always positive.
This immediately implies a difference
from the ordinary black hole entropy, computed above,
for the case of spin one fields~\cite{kabat}.
On the other hand, for scalar fields
in ordinary black hole
backgrounds both entropies are identical,
since in that case sign ambiguities in the entropy
do not arise. On the other hand, our computation for the
extreme EYMH case, provided the latter exists, has shown that, in general,
one should expect a difference between the two entropies
even in the case of scalar fields propagating
in such (extreme) non-Abelian black hole backgrounds.
\pr
There exists, of course,
the interesting possibility that the entanglement entropy of scalar
fields in this extreme black hole background
can be identified with the logarithmic entropy terms (\ref{entrop3}),
in which case the latter must be positive definite.
This, as we discussed above, would imply restrictions
on the boundary (horizon) values of the gauge hair for the extreme black hole
to exist. The restrictions seem to be relatively mild though.
As an example of the kind of the situation one
encounters in such cases, consider the case
where
extreme EYMH black hole solutions exist.
{}From (\ref{curious}),  we observe that positivity
of $m _h ''$ implies restrictions on the size of $\omega _h$,
$2\omega _h  >  \phi _h^2 ( 1 + \omega _h)^2 > 0$, which
is a mild restriction.
\pr
However, all these are mere speculations at this stage.
One has to await for
a complete analytic solution of the EYMH black hole problem
before reaches any conclusions regarding entropy production
and information loss in extreme cases.
Therefore, we leave any further considerations
on such issues for future work.

\section{Conclusions and Outlook}
\pr
In this work,
we have analyzed in detail black holes
in (3+1)-dimensional Einstein-Yang-Mills-Higgs
systems. We have argued that the conditions for
the no-hair theorem are violated, which allows
for the existence of Higgs and non-Abelian hair.
This analytic work supports the numerical evidence for the
existence of hair found in \cite{greene}.
This is due to a balance between the
gauge field repulsion and the gravitational attraction.
However we have shown that the above black holes
are unstable, and therefore cannot
be formed by gravitational collapse of stable matter.
Although the instability of the black hole
sphaleron sector
was expected for topological reasons, however
our analysis in this work, which includes an {\it exact}
counting of the unstable modes in this sector,
acquires value in the
sense that we have managed to
describe rigorously the sphaleron black holes
from a mathematical point of view.
In the gravitational sector we have
used catastrophe theory to classify
and count the unstable modes.
Our method of using as a catastrophe
functional the black hole mass and as
a control parameter the Higgs field v.e.v.
proved advantageous over existing methods
of similar origin~\cite{torii} in that
we managed to understand the
connection with the Schwarzschild black holes
from a stability/catastrophe-theoretic
point of view.
The above analysis, although applied
to a specific class of systems, however
is quite general and the various steps
can be applied to other self-gravitating structures
in order to reach conclusions related to
the existence of non-trivial hair
and their stability. For instance, we can tackle
the problem of moduli hair
of black holes in string-inspired dilaton-coupled
higher derivative gravity~\cite{kanti}.
The presence of Gauss-Bonnet combinations in such
systems shares many similarities with the case of
the non-Abelian black holes, and it would be
interesting to study in detail the possibility
of having non-trivial hair (to all orders in
the Regge slope $\alpha '$) and its stability,
following the methods advocated in the present work.
\pr
In addition to the question of the stability
of non-conventional hairy solutions,
the above analysis has revealed
another important aspect
concerning the information theoretic content
of these (3+1)-dimensional hairy black holes, namely the existence of
logarithmic divergent contributions to the entropy of
matter (quantum (scalar) fields) near the horizon.
Such contributions owe their existence
to the non-trivial hair of the black hole, and they modify
the Bekenstein-Hawking entropy formula, by yielding contributions
that do not depend on the horizon area.
Our findings can be compared to a similar
situation characterizing
extreme (string-inspired) black holes~\cite{dilatonbh}. There,
the deviation from the Bekenstein-Hawking entropy
was seen to occur by the fact that in the extreme case, due to the
presence of the dilaton, the effective horizon area vanishes, whilst
the entropy did not vanish. In our case, despite the non vanishing
entropy in the extreme case,  the logarithmically-divergent
entropy contributions violate explicitly the classical entropy-area
formula by yielding contributions that are independent of the horizon area.
This kind of entropy is clearly associated with loss of information
across the horizon but it is not described in terms
of classical geometric characteristics of the black hole.
If true in a full quantum theory of gravity,
this phenomenon might explain the information paradox.
The question of associating this entropy with the
entanglement entropy of fields in the EYMH background
is left open in the present work. We hope to come
back to this issue in the near future.
\pr
Whether a full quantum theory of gravity could
make sense of such divergencies or not remains to be seen.
There are conjectures/indications that string theory,
which is believed to be a mathematically
consistent, finite theory of quantum gravity, yields
finite extensive quantities at the horizon~\cite{susskind},
if string states, which in a generalized sense are gauged states,
are properly taken into account~\cite{emn}.
However, our understanding of these issues,
which are associated with the incompatibility - at present at least -
of canonical quantum gravity with quantum mechanics, is so incomplete
that any claim or attempt to relate the above issues to
realistic computations
involving quantum black hole physics would be inappropriate.
We think, however, that
it is interesting to point out yet another contradiction
of quantum mechanics and general relativity associated
with the proper quantization of extended objects possessing space-time
boundaries.

\pr
\nk {\Large {\bf  Acknowledgements}}
\pr
N.E.M. would like to thank the organizers of the {\it 5th Hellenic School
and Workshop
on Particle Physics and Quantum Gravity}, Corfu (Greece), 3-25 September
1995, for the opportunity they gave him to present
results of the present work. E.W. gratefully
acknowledges E.P.S.R.C. for a research studentship.
We also thank J. Bekenstein for a useful correspondence.
\newpage
\section*{Appendix A}
\subsection*{Notation and conventions}

Throughout this paper we use the sign conventions of Misner, Thorne
and Wheeler \cite{misner}
for the metric and curvature tensors. In particular, the signature of
the metric is $(-+++)$. For the EYMH system, we write the most general
spherically symmetric metric in the form
\be
ds^{2}= -NS^{2} dt^{2}+N^{-1} dr^{2} +r^{2}(d\theta ^{2}
+ \sin ^{2} \theta d\varphi ^{2})
\ee
where $N$ and $S$ are functions of $t$ and $r$ only and can be written
in terms of the mass function $m$ and the function $\delta $ as
\be
N(t,r)=1-\frac {2m(t,r)}{r}, \qquad
S(t,r)=e^{-\delta (t,r) }.
\ee
This latter form of the metric is particularly useful for black hole
space-times.
Following ref. \cite{greene}, we take the most general spherically symmetric
SU(2) gauge potential in the form
\be
A=a_{0} \tau_{r} dt + a_{1} \tau_{r} dr
+(1+\omega )[ \tau_{\theta } \sin \theta d\varphi -\tau _{\varphi}
d\theta ]
+{\tilde {\omega }} [\tau _{\theta } d\theta + \tau _{\varphi}
\sin \theta d\varphi ]
\ee
where $a_{0}$, $a_{1}$, $\omega $ and ${\tilde {\omega }}$ are
functions of $t$ and $r$ alone and the $\tau _{i}$ are given by
\bea
\tau _{r} & = &
\tau _{1} \sin \theta \cos \varphi +
\tau _{2} \sin \theta \sin \varphi +
\tau _{3} \cos \theta \\
\tau _{\theta } & = &
\tau _{1} \cos \theta \cos \varphi +
\tau _{2} \cos \theta \sin \varphi -
\tau _{3} \sin \theta \\
\tau _{\varphi } & = &
-\tau _{1} \sin \varphi +
\tau _{2} \cos \varphi
\eea
with $\tau _{i}$, $i=1,2,3$ the usual Pauli spin matrices.
The complex Higgs doublet assumes the form
\be
\Phi =\frac {1}{\sqrt {2}} \left(
\begin{array}{c}
\psi _{2}+i\psi _{1} \\
\phi -i\psi _{3}
\end{array}
\right)
\ee
where a suitable spherically symmetric ansatz is
\be
{\mbox {\boldmath $\psi $}}
=\psi (t,r) {\mbox {\boldmath ${\hat r}$}},
\qquad
\phi = \phi (t,r).
\ee
Then the EYMH Lagrangian is \cite{greene}
\bea
{\cal L}_{EYMH}  &=&
-\frac {1}{4\pi} \left[
\frac {1}{4} |F|^{2} +
\frac {1}{8} (\phi ^{2} +|\psi |^{2} )|A|^{2}
+\frac {1}{2} g^{MN}\left[ \partial _{M} \phi \partial _{N}\phi
+(\partial _{M} \psi ) \cdot (\partial _{N} \psi ) \right]
\right. \nn \\ & & \left.
+V(\phi ^{2}+ |\psi |^{2})
+\frac {1}{2} g^{MN} A_{M} \cdot [
\psi \times \partial _{N} \psi +
\psi \partial _{N} \phi -
\phi \partial _{N} \psi ]  \right]
\eea
and the Higgs potential is
\be
V(\phi ^{2})=\frac {\lambda }{4}  (\phi ^{2} -v^{2})^{2}.
\ee
\pr
For the equilibrium static solutions,
$a_{0}$, $a_{1}$, $\tilde \omega $ and $\psi$ all vanish and the
remaining functions depend on $r$ only.
The metric functions $m(r)$ and $\delta (r)$ are required by the
Einstein equations to satisfy the following, where $'=d/dr $:
\bea
m'(r) & = &
\frac {1}{2} \left[
\left( 1-\frac {2m(r)}{r} \right) (2\omega '^{2} +r^{2} \phi '^{2})
\right] \nn \\ & &
+ \frac {r^{2}}{2} \left[
\frac {(1-\omega ^{2})^{2}}{r^{4}} +
\frac {\phi ^{2}}{2r^{2}} (1+\omega )^{2}
+ \frac {\lambda }{2} (\phi ^{2} -v^{2})^{2}
\right] \label{first} \\
\delta '(r) & = &
-\frac {1}{r} (2\omega '^{2} +r^{2}\phi '^{2})
\eea
subject to the boundary conditions $m(r_{h})=\frac {r_{h}}{2}$
in order for a regular event horizon at $r=r_{h}$, and, in order for
the spacetime to be asymptotically flat, $\delta (\infty )=0$.
For an asymptotically flat spacetime, it is also the case that
$m(r) \rightarrow M$ as $r\rightarrow \infty $, where $M$ is a
constant equal to the ADM mass of the black hole.
Integrating (\ref{first}) from $r_{h}$ to $\infty $ we obtain:
\bea
{\cal M} -\frac {r_{h}}{2} & = &
m(\infty )-m(r_{h}) = \int _{r_{h}}^{\infty }
m'(r) dr \nn \\
 & = &
\int _{r_{h}}^{\infty } \frac {1}{2} \left[
\left( 1-\frac {2m}{r} \right) (2\omega '^{2} +r^{2} \phi '^{2})
\right] \nn \\ & &
+\frac {r^{2}}{2} \left[ \frac {(1-\omega ^{2})^{2}}{r^{4}}
+ \frac {\phi ^{2}}{2r^{2}} (1+\omega )^{2}
+\frac {\lambda }{2} (\phi ^{2}-v^{2})^{2} \right]
\eea
This equation defines the mass functional $\cal M$ as an integral of
the fields over the spacetime.
\pr
Finally we define the `tortoise' co-ordinate $r^{*}$ by
\be
\frac {dr^{*}}{dr} = \frac {1}{NS}.
\ee
\pr
 \subsection*{Numerical solution of equilibrium equations}
The static field equations for the metric functions and matter fields
are:
\bea
m'(r) & = & \frac {1}{2} \left[ \left( 1-\frac {2m}{r}\right)
(2\omega '^{2} +r^{2} \phi '^{2}) \right]
\nn
\\ & &
+\frac {r^{2}}{2} \left[ \frac {(1-\omega ^{2})^{2}}{r^{4}} +
\frac {\phi ^{2}}{2r^{2}} (1+\omega )^{2} +
\frac {\lambda }{2} (\phi ^{2}-v^{2})^{2} \right] \label{third} \\
\delta '(r) & = &
-\frac {1}{r} (2\omega '^{2} +r^{2}\phi '^{2})\label{second}  \\
N\omega '' & = &
-\frac {(NS)'}{S} \omega' + \frac {1}{r^{2}} (\omega ^{2}-1)\omega
+\frac {\phi ^{2}}{4} (1+\omega ) \\
N\phi '' & = &
-\frac {(NS)'}{S} \phi' -
\frac {2N}{r} \phi' +
\frac {\phi }{2r^{2}} (1+\omega )^{2} +
\lambda \phi (\phi ^{2}-v^{2}) \label{fourth}
\eea
For finite energy solutions, we require that $\omega (\infty )=-1$,
$\phi (\infty )=v$ and $\delta (\infty )=0$ in order for spacetime
to be asymptotically flat.
These equations trivially possess the Reissner-Nordstr\"{o}m solution
given by
\be
m\equiv \frac {r_{h}}{2}, \qquad
\omega \equiv -1, \qquad
\phi \equiv v, \qquad
\delta \equiv 0.
\ee
Non-trivial solutions do not occur in closed form, so a numerical
method of solution is necessary as in \cite{greene}.
We set the horizon radius $r_{h}=1$ and $\lambda =0.15$
(cf. $\lambda =0.125$ in  ref. \cite{greene}).
\pr
{}From the above equations, if the function $\delta (r)$ satisfies
(\ref{second}) then $\delta (r) + \mbox { constant }$ will also be a  valid
solution.
To make the numerical solution easier, we set $\delta (r_{h})=0$
(so that $\delta (\infty )=0$  will not be satisfied) when
integrating outwards from $r_{h}$. An appropriate constant can then
be added to $\delta (r)$, after the field equations have been solved,
to ensure that the boundary condition at infinity holds.
\pr
With this transformation, there are two unknowns at the event horizon,
 $\omega (r_{h})$ and $\phi (r_{h})$, since the field equations yield
\bea
\omega '(1) & = &
\frac {\frac {1}{4} \phi ^{2}_{h} (1+\omega _{h})
-\omega _{h}(1-\omega _{h}^{2}) }
{1-(1-\omega _{h}^{2})^{2}-\frac {1}{2}\phi ^{2}_{h}(1+\omega _{h})^{2}
-\frac {\lambda }{2}(\phi ^{2}_{h}-v^{2})^{2}} \nn \\
\phi '(1) & = &
\frac {\frac {1}{2} \phi _{h} (1+\omega _{h})^{2}
+\lambda \phi _{h} (\phi _{h}^{2}-v^{2})}
{1-(1-\omega _{h}^{2})^{2}-\frac {1}{2}\phi ^{2}_{h}(1+\omega
_{h})^{2}
-\frac {\lambda }{2}(\phi ^{2}_{h}-v^{2})^{2}}
\label{syst}
\eea
where
\be
\omega _{h}=\omega (r_{h})=\omega (1)  \qquad
\phi _{h}=\phi (r_{h})=\phi (1).
\ee
Solving the field equations (\ref{third})--(\ref{fourth})
is  therefore a two-parameter
shooting problem.
The procedure is to take initial `guesses' for the unknowns
$\omega _{h}$ and $\phi _{h}$ and then integrate the differential
equations out from $r_{h}$ using a standard ordinary differential
equation solver, attempting to satisfy the boundary conditions for
large $r$.
The initial starting values for $\omega _{h}$ and $\phi _{h}$
are then adjusted until these boundary conditions are satisfied
(see \cite{press} for further details of the algorithm used).
\pr
For each fixed value of the Higgs mass $v$, there are many solutions
which can be indexed by the number of nodes $k$ of the potential
function $\omega (r)$.
Here we concentrate on the case $k=1$.
Then, for each $v$, there are two solutions which can be ascribed
to one of two families of solutions: the $k=1$ branch or the
quasi-$k=0$ branch, depending on the behaviour of the families as
$v\rightarrow 0$.
The quasi-$k=0$ branch of solutions approaches the Schwarzschild
solution $\omega \equiv 1$, $\phi \equiv 0$ as $v\rightarrow 0$,
whereas the $k=1$ branch of solutions approaches the first coloured
black hole of \cite{bizon} as $v\rightarrow 0$, with
$\phi \equiv 0$.
As $v$ increases, the two branches of solutions join up at
$v=v_{max}=0.352$.
This phenomenon does not occur for $\lambda =0.125$, as found
by Greene, Mathur and O'Neill \cite{greene}.
However, they conjectured that the two branches of solutions would
converge of some value of $\lambda $.
We stress here that our approach is somewhat different from that
of ref. \cite{torii}, where the field equations were solved for
fixed Higgs mass $v$ and varying $r_{h}$, whereas we have fixed
$r_{h}$ and varied $v$.
\pr
For each value of the Higgs mass $v$, we calculated the quantities
\bea
{\cal M} & = &
\frac {r_{h}}{2} + \int _{r_{h}}^{\infty } \left\{
 \frac {1}{2} \left[ \left(
1-\frac {2m}{r} \right) (2\omega '^{2} +r^{2} \phi '^{2})
\right] \right.   \nn \\ & &
+r^{2}\left.  \left[
\frac {(1-\omega ^{2})^{2}}{r^{4}} +
\frac {\phi ^{2}}{2r^{2}}(1+\omega )^{2} +
\frac {\lambda }{2} (\phi ^{2}-v^{2})^{2} \right] \right\}  dr
\\
\delta _{0} & = &
\int _{r_{h}}^{\infty} \frac {1}{r} (2\omega '^{2}+
r^{2}\phi '^{2}) dr
\eea
for each of the two solutions.
The resulting solution curve plotted in $(v,\delta _{0},{\cal M})$
space is shown in figure 1.
The projection of this curve on to the $(v,{\cal M})$
 plane are shown in
figure 2.
\pr
One issue that is important, especially when we come to consider
the thermodynamics and entropy of the black holes, is whether or
not they are extremal.
An extremal black hole occurs when $N$ has a double zero at the
event horizon, and is caused physically by an inner horizon moving
outwards until it coincides with the outermost event horizon.
Mathematically, the condition for extremality is that
\be
m'(1) = \frac {1}{2}.
\label{extrem2}
\ee
{}From the field equations (\ref{fourth}), we have
\be
m'(1) =\frac {1}{2} \left[
(1-\omega _{h}^{2})^{2} + \frac {1}{2} \phi _{h}
(1+\omega _{h})^{2} +\frac {\lambda }{2}
(\phi _{h} ^{2}-v^{2})^{2} \right]
\label{extrem}
\ee
\be
m''(r_{h})=\frac {1}{r_{h}} \left( 1-
\frac{\phi _{h}^{2}}{2\omega _{h}} (1+\omega _{h})^{2} \right)
\label{double}
\ee
where in the last relation we have kept an explicit $r_{h}$
dependence for calculational convenience.
There is thus no {\it a priori} reason why this quantity should not be
equal to one half for some equilibrium solution.
For the solutions on the $k=1$ and quasi-$k=0$ branches, we can
however place the following bounds on $m'(1)$.
The first term is decreasing for $\omega _{h}$ positive and
increasing,
and hence is bouded above by its value for the smallest value of
$\omega _{h}$ along these branches, which is $\omega _{h}=0.632$, whence
\be
(1-\omega _{h}^{2})^{2} \le 0.360.
\ee
Along both these branches, $\phi _{h} \le 0.19 v$ which gives the
following bound on the second term,
\be
\frac {1}{2} \phi _{h}^{2}(1+\omega _{h})^{2} \le
2 \times (0.19v)^{2} \le 2 \times 0.19^{2} \times 0.352^{2}
= 8.95 \times 10^{-3}.
\ee
Finally, for the last term we have
\be
\frac {\lambda }{2} (\phi _{h}^{2}-v^{2})^{2}
\le \frac {\lambda }{2}v^{4} \le 0.15 \times 0.5 \times 0.352^{4}
= 1.15 \times 10^{-3}.
\ee
Adding together all the contributions, we find that
\be
m'(1)\le 0.5 \times ( 0.360+8.95\times 10^{-3} + 1.15\times 10^{-3} )
=0.185
\le 0.5
\ee
and hence all the equilibrium black holes considered here are
non-extremal.
\pr
\subsection*{Linear perturbation equations}
Consider small, time-dependent perturbations about the equilibrium
solutions discussed above, within the initial ansatz for the
metric and matter field functions.
We use a $\delta $ to denote one of these small perturbation
quantities, all other quantities are assumed to be static equilibrium
functions.
Following ref. \cite{bs}, we set $\delta a_{0}=0$ so that the
field configurations remain purely magnetic.
With this choice, the perturbation equations decouple into two
independent coupled systems.
The first concerns $\delta a_{1} $, $\delta {\tilde {\omega }}$
and $\delta \psi $ only.
The equations take the form, with a prime denoting $d/dr^{*}$
where $r^{*}$ is the tortoise co-ordinate:
\bea
-Nr^{2}{\ddot {\delta a_{1}}} & = &
2N^{2}S^{2} \left( \omega ^{2} +\frac {r^{2}}{8} \phi ^{2}\right)
\delta a_{1} + 2NS (\omega  \delta {\tilde \omega }'
-\omega '\delta {\tilde {\omega }}) \nn \\
 & &
+\frac {1}{2} r^{2} NS (\phi ' \delta \psi -\phi \delta \psi ') \\
2{\delta {\ddot {\tilde {\omega }}}} & = &
2(NS\omega \delta a_{1})' +
2NS\omega ' \delta a_{1} +\delta {\tilde {\omega }}'' +
NS^{2}\phi \delta \psi  \nn \\
 & &
-\frac {2}{r^{2}}S^{2} \left( \omega ^{2}-1 +\frac {\phi ^{2}}{4}
\right) \delta {\tilde {\omega }} \\
-r^{2} \delta {\ddot {\psi }} & = &
\frac {1}{2} (NSr^{2} \phi \delta a_{1})'
+ \frac {1}{2} r^{2} NS \phi ' \delta a_{1}
-NS^{2} \phi \delta {\tilde {\omega }} -(r^{2}\delta \psi ')' \nn \\
 & &
+2NS^{2} \left( \frac {(1-\omega )^{2}}{4}+\frac {1}{2} r^{2}
\lambda (\phi ^{2} -v^{2}) \right) \delta \psi \\
0 & = & \partial _{t}
\left\{ \left( \frac {r^{2}}{S} \delta a_{1} \right) '
+ 2\omega {\tilde {\omega }} -\frac {r^{2}}{2} \phi \delta \psi
\right\}.
\eea
This final equation is known as the {\it Gauss constraint} equation,
since it represents an additional constraint on the field
perturbations rather than an equation of motion.
This system of coupled equations is referred to as the
{\it sphaleronic sector} because it does not involve any perturbations
of the metric functions.
\pr
The remaining perturbation equations form the {\it gravitational
sector} and concern the perturbations of the metric functions and also
$\delta \omega $ and $\delta \phi $:
\bea
-\delta {\ddot {\omega }}  & = &
-\delta \omega '' +U_{\omega \omega } \delta \omega
+U_{\omega \phi }\delta \phi  \\
-\delta {\ddot {\phi }} & = &
-\delta \phi '' +U_{\phi \omega } \delta \omega
+U_{\phi \phi }\delta \phi
\eea
where the $U$'s are complicated functions of $N$, $S$, $\omega $ and
$\phi $ and are given explicitly   in section 4, equation \ref{4-2}.
The equations governing the behaviour of $\delta m$ and $\delta S$
are derived from the linearised Einstein equations and are:
\bea
\frac {d}{dr}(S\delta m ) & = & \frac {d}{dr} \left(
2NS \frac {d\omega }{dr} \delta \omega +r^{2}NS \frac {d\phi }{dr}
\delta \phi  \right)
\label{mreqn}  \\
\delta {\dot {m}} & = &
2N \frac {d\omega }{dr} \delta {\dot {\omega }} +
r^{2}N \frac {d\phi }{dr} \delta {\dot {\phi }}
\label{mteqn} \\
\delta \left( \frac {1}{S} \frac {dS}{dr} \right) & = &
\frac {4}{r} \frac {d\omega }{dr} \frac {d\delta \omega }{dr}
+2r \frac {d\phi }{dr} \frac {d\delta \phi }{dr}.
\label{seqn}
\eea
{}From (\ref{mreqn}) $ \delta m$ has the form
\be
\delta m =2N\frac {d\omega }{dr} \delta \omega +
Nr^{2} \frac {d\phi }{dr} \delta \phi + \frac {f(t)}{S}
\label{compare}
\ee
where $f(t)$ is an arbitrary function of $t$.
Compare this with the following, which results from integrating
(\ref{mteqn}):
\be
\delta m = 2N\frac {d\omega }{dr} \delta \omega +
Nr^{2} \frac {d\phi }{dr} \delta \phi + g(r)
\label{compare1}
\ee
where $g(r)$ is an arbitrary function of $r$.
Comparing (\ref{compare}) with (\ref{compare1}), we see that
$f(t) \equiv 0 \equiv g(r) $ and
\be
\delta m = 2N\frac {d\omega }{dr} \delta \omega +
Nr^{2} \frac {d\phi }{dr} \delta \phi .
\ee
\pr
We consider periodic perturbations of the form
\be
\delta \omega (r,t) =\delta \omega (r) e^{i\sigma t}
\ee
and similarly for the other pertubation quantities.
When substituted into the perturbation equations for each of the two
sectors, the equations studied in detail in sections 3 and 4 are
derived.
\pr

\section*{Appendix B}
\subsection*{Definitions and results of catastrophe theory}
Consider a family of functions
\be
f:X \times C \rightarrow  \BbbR
\qquad f(x,c)=f_{c}(x)
\ee
Here $X$ and $C$ are both manifolds known as the state space and
control space respectively.
In other words, we have a family of functions of the variable $x$,
the members of the family being indexed by $c$.
{}From now on we take both $X$ and $C$ to be intervals of the real line.
Then $f$ maps out a surface $z=f(x,c)$ in $\BbbR ^{3}$
which is known as the {\it Whitney surface }\cite{poston}.
\pr
The catastrophe manifold is defined as the subset of $X\times C$
at which
\be
\frac {d}{dx} f_{c}(x) =0,
\ee
namely it is the set of all critical points of the family of
functions. In section 4, critical points of the functional
$\cal M$ correspond to solutions of the field equations, and
hence the catastrophe manifold corresponds to the projection of
the solution curve onto the $(x,c)=(\delta _{0},v)$ plane.
\pr
The catastrophe map $\chi $ is the restriction to the catastrophe
manifold of the natural projection
\be
\pi : X \times C \rightarrow C,
\qquad
\pi (x,c)=c.
\ee
This can easily be extended to a projection of the solution curve
on to the $(c,z)$ plane:
\be
\chi (x,c,z=f(x,c)) =(c,z=f(x,c)).
\ee
The singularity set is the set of singular points of $\chi $
in the catastrophe manifold, and the image of the singularity
set in $C$ is called the {\it bifurcation set }$B$.
Here both manifolds $X$ and $C$ are of dimension 1, and hence $\chi $
will be singular whenever its derivative vanishes.
\pr
The first result we require is that the singularity set  is the
set of points $(x,c)$ at which $f_{c}(x)$ has a degenerate
critical point, in other words, both
\be
\frac {d}{dx} f_{c}(x) =0 {\mbox { and }}
\frac {d^{2}}{dx^{2}} f_{c}(x) =0.
\ee
This implies that the set $B$ is the place where the number and
nature of the critical points of the family of functions $f_{c}(x)$
change (see \cite{poston} for more details of these results).
\pr
In our case, where both $X$ and $C$ are one-dimensional,
the only possibility is that the bifurcation set $B$ either is
empty (in which case there is no catastrophe) or $B$ contains
a single point (when a {\it fold catastrophe } occurs).
We observe in section 4 that the latter situation arises.
\pr
The catastrophe manifold is a curve $\cal C$ in the $(x,c)$ plane,
with the point $B$ lying on it.
On one side of the point $B$, points lying on $\cal C$ correspond
to maxima of the functions $f_{c}(x)$ whilst on the other side of $B$
they represent minima.
In section 4 the value of $f$ corresponds to energy, so that minima of
$f$ will represent (relatively) stable objects, whilst maxima of $f$
will represent (relatively) unstable configurations.

\newpage
{\Large {\bf Figure Captions }}
\pr
\nk {\bf Figure 1} Solution curve for black holes in EYMH theory
with one node of the gauge field component $\omega $, in
$(v,\delta_0, {\cal M})$ parameter space:  $v$ is the Higgs v.e.v.,
$\delta _0$ is the black hole parameter defined in Appendix A and
${\cal M}$ is the mass functional of the black hole. Notice that
the solution curve is smooth, but this is not true for its projection
onto the $(v, {\cal M})$ plane, see Figure 2 below.
\pr
\pr
\nk {\bf Figure 2} Projection of the solution curve of Figure 1
onto the $(v, {\cal M})$ plane. The cusp at $v=v_{max}$ indicates
the r\^ole of ${\cal M}$ as a fold catastrophe functional, with $v$
the appropriate control parameter. The upper branch of solutions
(quasi-$k=0$),
corresponding to higher entropy, is more stable relative to the lower
branch ($k=1$).

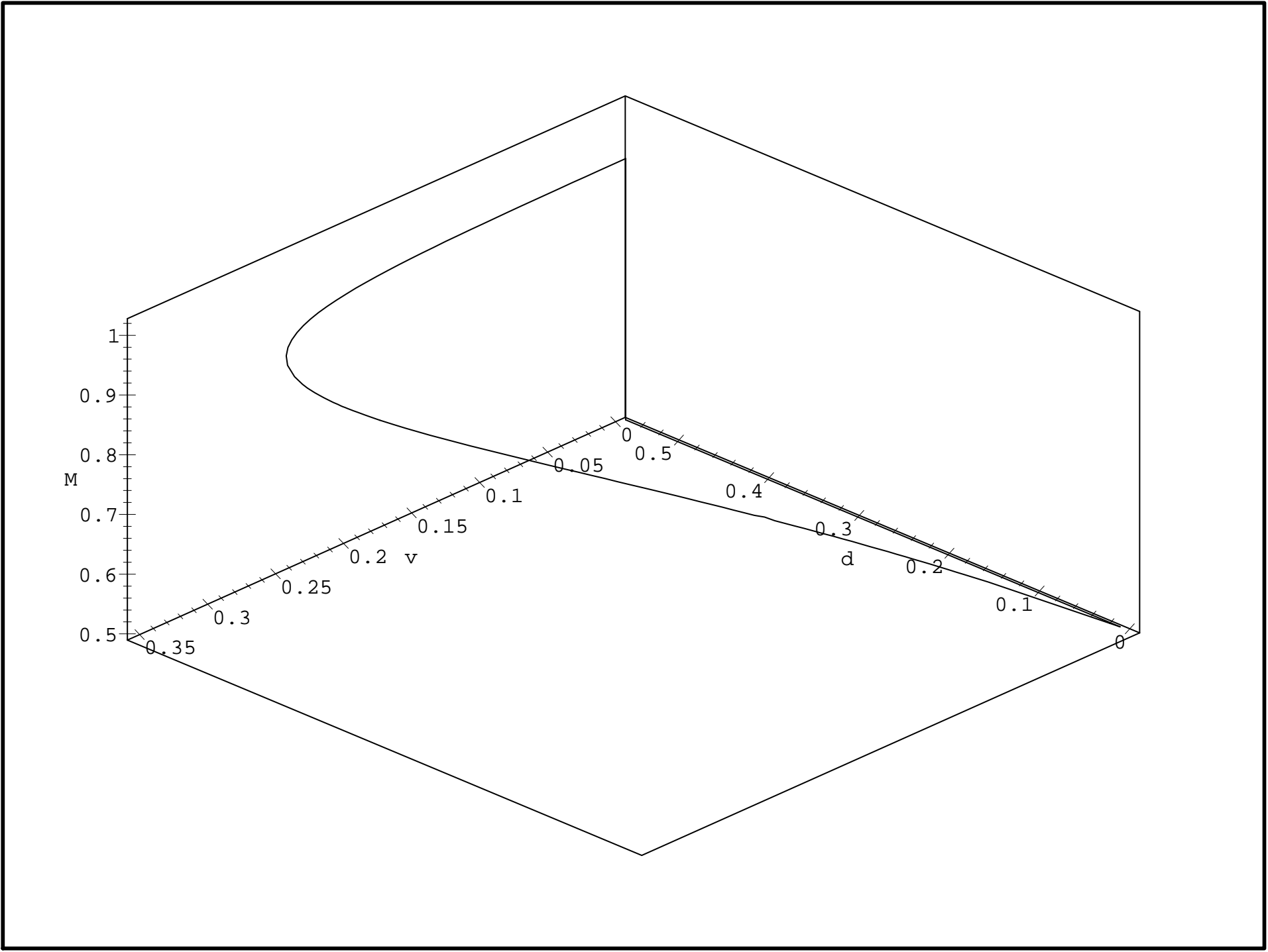
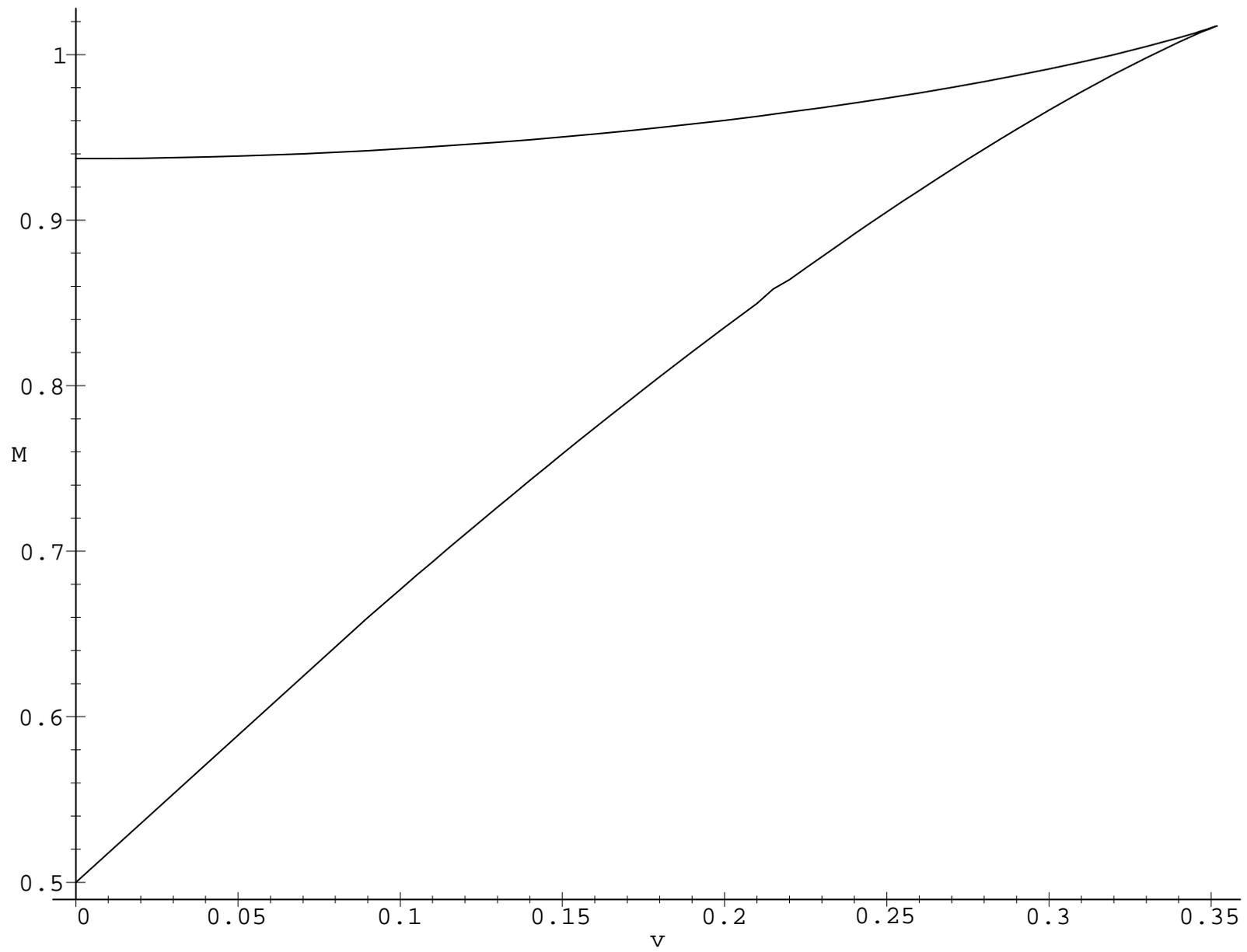
\end{document}